\documentclass[11pt,a4paper]{article}
\pdfoutput=1
\usepackage{jheppub}	

\usepackage{graphicx}

\def\lsim{\raise0.3ex\hbox{$\;<$\kern-0.75em\raise-1.1ex
\hbox{$\sim\;$}}}
\def\gsim{\raise0.3ex\hbox{$\;>$\kern-0.75em\raise-1.1ex
\hbox{$\sim\;$}}}

\def\be{\begin{equation}}
\def\ee{\end{equation}}
\def\ba{\begin{eqnarray}}
\def\ea{\end{eqnarray}}

\title{Does $H \to \gamma \gamma$ Taste Like Vanilla New Physics?}
\author[a]{L.~G.~Almeida}
\author[a]{E.~Bertuzzo}
\author[a,b]{P.~A.~N.~Machado}
\author[a,b]{R.~Zukanovich Funchal} 

\emailAdd{leandro.almeida@cea.fr}
\emailAdd{enrico.bertuzzo@cea.fr}
\emailAdd{accioly@fma.if.usp.br}
\emailAdd{zukanov@if.usp.br} 

\affiliation[a]{Institut de Physique Th\'eorique, CEA-Saclay, 91191 Gif-sur-Yvette, France}
 \affiliation[b]{
Instituto de F\'{\i}sica, Universidade de S\~ao Paulo, 
 C.\ P.\ 66.318, 05315-970 S\~ao Paulo, Brazil}

\abstract{ 
We analyse the interplay between the Higgs to diphoton rate and 
electroweak precision measurements constraints in extensions of 
the Standard Model with new uncolored charged 
fermions that do not mix with the ordinary ones. We also compute the 
pair production cross sections for the lightest fermion and compare 
them with current bounds. 
} 

\arxivnumber{1207.5254}

\begin{document}
\maketitle

\section{Introduction}
\label{sec:intro}

The Standard Model (SM) has been so far exhaustedly ratified by
experiments.  The recent discovery of a new boson, compatible with the
SM Higgs particle, reported by ATLAS~\cite{ATLAS} and CMS~\cite{CMS}
experiments at the Large Hadron Collider (LHC), finally inaugurates a
new era in the field. For the first time we have access to the
electroweak symmetry breaking sector of the SM and to the mysteries it
may reveal.

Indeed, the combined analyses~\cite{Corbett:2012dm,Giardino:2012dp,
  Espinosa:2012im} of ATLAS and CMS results at 7 and 8
TeV~\cite{atlas7new,Chatrchyan:2012tx, cmspashig12015,cmspashig12020,
  atlasconf12091} with the Tevatron experiments data~\cite{Tevatron}, 
seem to suggest that physics beyond the SM is already here. There are
two important facts that all these analyses seem to indicate: the
gluon-gluon Higgs production cross section is $\sim$ 40\% smaller and
the Higgs to diphoton decay $\sim$ 3 larger than the SM
prediction. Furthermore, the CMS uncombined Higgs data~\cite{CMS}
clearly corroborate to this assessment. Their $H \to \gamma \gamma$
events tagged as produced by vector boson fusion prefer a production
cross section $\sim 2$ times higher than the SM value, while their $H
\to \gamma \gamma$ untagged events prefer it only $\sim 50\%$
higher. The former can be explained by an increase of $\Gamma(H\to
\gamma \gamma)$ whereas the latter by an additional decrease of the
gluon-gluon Higgs production cross section. Their $H \to Z Z$ events
also point towards a $\sim 30\%$ lower gluon-gluon cross section.

In this paper we focus on the Higgs to diphoton decay as a possible
smoking gun to new physics.  The $H\to \gamma \gamma$ partial decay
width, which appear only at loop level, is sensitive to the existence
of extra charged states that couple to the Higgs boson.  Such states
arise in a variety of beyond the SM models. Thus, since the first
measurements by the LHC experiments suggested the observed Higgs to
diphoton rate was larger than the SM
one~\cite{Chatrchyan:2012tw,ATLAS:2012ad}, a number of works have
studied the effect of new particles in this
width~\cite{Cao:2011pg,Alves:2011kc,Carena:2011aa,Draper:2012xt,Kumar:2012ww,Dawson:2012di,Carena:2012gp,Akeroyd:2012ms,Carmi:2012zd,Carena:2012xa,Carmi:2012in,Chang:2012ta,Bertolini:2012gu}.

We propose here to map, in a model independent fashion, the properties
(masses, charges, couplings) that the new particles have to satisfy in
order to account for the observed $H\to \gamma \gamma$ width in an
economic way and pass the electroweak precision tests. We consider the
effect of extra fermions, in their smallest allowed $ \rm SU(2)_L
\times U(1)_Y$ representations, assuming that possible additional
states, that could accompany these fermions in a complete model,
are sufficiently heavy to play no significant role. We also compute,
in each case, their production cross section at the LHC and discuss
possible signatures and limits.

This work is organized as follows. In the Sec.~\ref{sec:model} we
present the general framework of our approach to the $H \to \gamma
\gamma$ width and in Sec.~\ref{sec:clfermions} we discuss the
doublet-singlet and triplet-doublet uncolored fermion states and their
corresponding lagrangians that we introduced in order to increase the
diphoton rate (the specific choice $y=1/2$ for the doublet hypercharge
in the triplet-doublet model corresponds to the supersymmetric
Wino-Higgsino case). In Sec.~\ref{sec:hgg} we discuss what kind of
enhancements are possible with these models and their dependency on
the model parameters.  In Sec.~\ref{sec:STU} we examine the
constraints on the parameter space from the electroweak precision
tests. In Sec.~\ref{sec:production} we consider the predictions for
the pair production cross sections of these new fermions at the LHC
operating at a center of mass energy of 8 TeV.

\section{General Framework}
\label{sec:model}

We assume the new 125 GeV particle observed at the LHC is in fact a
SM-like Higgs boson, responsible for the electroweak symmetry
breaking. It is a fundamental scalar transforming as part of the $\rm SU(2)_L$
doublet 

\begin{equation}
H = \left( \begin{array}{c}  h^+ \\ h^0 \end{array} \right) , 
\end{equation}
with the SM Higgs charge assignments and hypercharge Y=1/2.

The new particles will not mix with the SM fermions, they will only
couple to the Higgs and the gauge sector respecting the SM symmetry
group.  This is feasible in a concrete model by introducing a new
quantum number in connection to an unbroken or nearly unbroken
symmetry, exclusive to the new sector.  We will consider colorless
fermion states in their lowest allowed $\rm SU(2)_L \times U(1)_Y$
representations, \emph{i.e.}  singlets, doublets and triplets. Once the
representation is chosen, their couplings with the SM gauge boson will
be basically fixed. The only free parameters will be their couplings
to the Higgs, their charges and their masses.  We do not study here
particles with $\rm SU(3)_C$ quantum numbers, for simplicity and
because we are not interested in this work to change the Higgs
production cross section.

We will examine the allowed regions of these parameters in order 
for these new particles to significantly contribute to the 
Higgs diphoton width. We will do this by imposing 
$1.4 < \Gamma (H \to \gamma \gamma)/\Gamma^{\rm SM}(H\to \gamma \gamma)
<5.4$ at 95\% CL~\cite{Corbett:2012dm}.

The Higgs to diphoton decay can be written in terms of the couplings 
to the particles in the loop as

\begin{equation}
\Gamma (H \to \gamma \gamma) = \displaystyle \frac{\alpha^2\,
  m_H^3}{1024 \pi^3} \left \vert \frac{2}{v}A_{1} (\tau_W) + \frac{8}{3v}
A_{1/2}(\tau_t) + \frac{2 g_{H f \bar f}}{m_f} N_{c,f}
q_f^2 A_{1/2}(\tau_f) + \frac{g_{HSS}}{m_S^2} N_{c,S} q_S^2
A_{0}(\tau_S) \right \vert^2  \, , 
\label{eq:hgg}
\end{equation}
where $\tau_a \equiv (m_H/2 m_a)^2$, $a=W,t,f,S$, $m_H$ is the Higgs
mass, $f$ ($S$) is a generic new fermion (scalar) with electric charge
$q_f$ ($q_S$), in units of the electric charge $e$, number of colors
$N_{c,f}$ ($N_{c,S}$) and mass $m_f$ ($m_S$), coupling to the Higgs
with strength $g_{Hf\bar{f}}$ ($g_{HSS}$). The loop functions $A_{1},
A_{1/2}$ and $A_0$ are defined in the Appendix.

The first and second contributions are the 
dominant SM ones, while the others are possible contributions
from extra fermions and scalars. Since for the W boson contribution
$A_1(\tau_W) \to -8.3$ and for the top quark $A_{1/2}(\tau_t) \to
+1.8$, to increase $H\to \gamma \gamma$ we need to include a new
negative contribution, 
comparable to the top one.

It was shown in Ref.\cite{Carena:2012xa}, where the leading contributions from 
new particles to the diphoton decay width was derived from the QED beta 
functions, that for fermions carrying the same electric charge and 
described by the mass matrix $M_{f}$ 
\begin{equation}
\frac{2 g_{H f_i \bar f_i}}{m_{f_i}} = \displaystyle \frac{\partial}{\partial v} \log \lambda^2_{f_i}(v),
\label{eq:ghff}
\end{equation}
where $\lambda^2_{f_i}(v)$ is an eigenvalue of $M^\dagger_f M_f$.
Clearly if fermions  cannot mix they will all contribute to the loop with
the same sign of the top contribution and decrease the Higgs to
diphoton width. So a required condition to enhance the diphoton
coupling to the Higgs is to have mixture. In this case the off-diagonal 
elements can enter with a term carrying the same sign of the W contribution, 
cancelling the top and increasing the width.

However, any physics beyond the SM must face its tremendous success:
fulfill the electroweak precision tests and evade direct detection bounds.

New states will inevitably contribute to the vacuum polarization
amplitudes of the electroweak gauge bosons $\Pi^{\mu \nu}_{ab}(q^2) =
- i g^{\mu \nu} \Pi_{ab}(q^2) + q^\mu q^\nu$
terms~\cite{Altarelli:1990zd,Peskin:1991sw}.
These new physics effects can be 
parametrized by the so-called quantum oblique parameters $S$, $T$ and 
$U$ defined as~\cite{Peskin:1991sw}

\begin{eqnarray}
\alpha(M_Z^2) \, S^{\text{NP}} & = & \frac{4 s_W^2 c_W^2}{M_Z^2} \left [\Pi^{\text{NP}}_{ZZ} (M_Z^2) - \Pi^{\text{NP}}_{ZZ} (0)
    -\Pi^{\text{NP}}_{\gamma \gamma}(M_Z^2) - \frac{c_W^2-s_W^2}{c_W
        s_W} \, \Pi^{\text{NP}}_{\gamma Z}(M_Z^2)\right] \nonumber
      \\ 
\alpha(M_Z^2) \, T^{\text{NP}} & = & \frac{\Pi^{\text{NP}}_{WW}(0)}{M_W^2}
        - \frac{\Pi^{\text{NP}}_{ZZ}(0)}{M_Z^2} \nonumber
          \\ 
\alpha(M_Z^2) \, U^{\text{NP}} & = & 4 s_W^2 \left [
            \frac{\Pi^{\text{NP}}_{WW}(M_W^2)-\Pi^{\text{NP}}_{WW}(0)}
                {M_W^2} - c_W^2 \left(
                \frac{\Pi^{\text{NP}}_{ZZ}(M_Z^2)-\Pi^{\text{NP}}_{ZZ}(0)}{M_Z^2}\right)
                    \nonumber \right .\\ & & \left . - 2 s_W c_W \,
                    \frac{\Pi^{\text{NP}}_{\gamma Z}(M_Z^2)}{M_Z^2} -
                       s_W^2 \, \frac{\Pi^{\text{NP}}_{\gamma
                            \gamma}(M_Z^2)}{M_Z^2} \right ],
\label{eq:STU}
\end{eqnarray}
where $s_W^2 = \sin^2\theta_W = 1- c_W^2 \equiv 1- M_W^2/M_Z^2$ , 
$M_Z$ and $M_W$ are, respectively, the Z boson and W boson masses. 
By  comparing the measurable electroweak observables with the 
theory prediction  one finds the fitted values~\cite{Baak:2011ze}

\begin{eqnarray}
\Delta S & = & S - S_{\rm SM}  =  0.04 \pm 0.10 \nonumber \\
\Delta T & = & T - T_{\rm SM}  =  0.05 \pm 0.11 \nonumber \\
\Delta U & = & U - U_{\rm SM}  =  0.08 \pm 0.11 
\label{eq:stulimits}
\end{eqnarray}
for the reference Higgs and top masses $M_{H,\rm ref} = 120$ GeV 
and $m_{t,\rm ref}= 173$ GeV, with the associated correlation matrix 

\begin{equation}
V= \left( \begin{array}{ccc}
1 & + 0.89 & -0.45 \\
+0.89 & 1 & -0.69 \\
-0.45 & -0.69 & 1
\end{array}\right).
\label{eq:corr}
\end{equation}


We will include these constraints in our models by minimizing the $\chi^2$ 
function defined as 

\begin{equation}
\chi^2 = \sum_{i,j} (X_i^{\text{NP}} -
X_i)(\sigma^2)^{-1}_{ij}(X_j^{\text{NP}} - X_j),
\label{eq:chi2}
\end{equation}
where $X_i= \Delta S, \Delta T, \Delta U$, are the fitted values of
the oblique parameters with their corresponding uncertainties
$\sigma_i$ defined in Eq.(\ref{eq:stulimits}), $X_i^{\text{NP}} =
S^{\text{NP}}, T^{\text{NP}}, U^{\text{NP}}$ are the contributions
from the extra states that we will be introduced in each model
investigated and $\sigma^2_{i,j} \equiv \sigma_i V_{ij}\sigma_j$.  We
will allow the values of the parameters of our models to vary such
that $\Delta \chi^2 = (3.53, 7.81, 11.3)$, which correspond to (68\%,
95\%, 99\%) CL in a three-parameter fit.  Since the difference between
$M_{H,\rm ref}$ and the actual Higgs mass $M_H=125$ GeV is rather
small and the uncertainties in the fitted parameters large, we will
not correct for the exact result of the Higgs contribution to the
oblique parameters.

Finally, since the new fermions couple to $Z$ and $\gamma$ they can be pair-produced at the LHC. We will examine, in
each case, the production cross-section and comment on possible
existing limits and perspectives. In order to calculate the production
cross sections we have implemented our models in
CalcHEP~\cite{calchep}.

\section{New Fermion States}
\label{sec:clfermions}
The existence of a chiral 4th generation that couples to the Higgs
boson is excluded by data, since heavy quarks would contribute to the
Higgs production cross section, increasing its rate by a factor $\sim$
9, and would exclude the Higgs up to $600$ GeV~\cite{CMS4th}.  To
avoid this problem our fermions will be vector-like.  We will also
assume that our fermions will have some new quantum number that
forbids mixing with the usual SM fermions. In this case, we need to
introduce at least two extra fermion fields in order to be able to
build a renormalizable coupling term with the SM Higgs field and to
have mixing. We will examine here the two smallest representations, {
  see Tab.~\ref{tab1}. Let us stress that the triplet-doublet case
  with $y=1/2$ corresponds to the supersymmetric Wino-Higgsino case.

\begin{table}[h!]\centering
\begin{tabular}{|c|c|c|c|}
\hline
  &\multicolumn{2}{|c|}{$\rm SU(2)_L$} & \\
\hline
Field & doublet-singlet & triplet-doublet & $\rm U(1)_Y$ \\
\hline\hline
$\chi_{L,R}$ & 2 & 3 & $\hat y=y-\frac{1}{2}$ \\
$\psi_{L,R}$ & 1 & 2 & $y$ \\
\hline
\hline
\end{tabular}
\caption{\label{tab1} Representations of the new fermions 
 and their corresponding hypercharges for the two cases we consider 
in this work.}
\end{table} 

\subsection{\bf Doublet-singlet model}

 The lagrangian describing the new fermion masses and couplings 
with the Higgs  is 

\begin{equation}
{-\cal L_{\rm H}^{\rm 2+1}} = c \, \overline{\psi_R} \, H \chi_L + c
\,\tilde{H} \, \overline{\chi_R} P_L  \psi_L + m_1 \overline{\chi_R} P_L \chi_L +
m_2 \overline{\psi_R} P_L \psi_L  + \rm h.c.,
\label{eq:h-2+1}
\end{equation}
where $\tilde{H}= i \tau_2 H^{*}$, $c$ is the Yukawa coupling to the 
Higgs, $P_{L,R} = \frac{1}{2}(1\pm \gamma_5)$,
$m_{1,2}$ are the vector-like $\chi,\psi$ masses.

After electroweak symmetry breaking, the Higgs acquires a vacuum expectation 
value (vev) $v$ endowing an extra mass contribution to the new fermions. 
The new fermions mass matrix takes the form
\begin{equation}
M_{2+1} = (\bar \psi_R \, \bar \chi_R^u\, 
\bar \chi_R^d) \left( \begin{array}{ccc}
m_2 & c v & 0 \\
c v &  m_1 & 0 \\
0 & 0 & m_1
\end{array}
\right) 
\left( \begin{array}{c}
\psi_L\\
\chi_L^u \\
\chi_L^d
 \end{array}\right)\, ,
\label{eq:mass-2+1}
\end{equation}
where we explicitly write the vector doublet as
\begin{equation}
\chi = \begin{pmatrix}
\chi^u \cr \chi^d
\end{pmatrix}
\end{equation}

To diagonalize $M_{2+1}$ we introduce the following transformations  
\begin{equation}
\omega_{L,R} \equiv
\left( \begin{array}{c}
\omega^{1}_{L,R}\\
\omega^{2}_{L,R}
 \end{array}\right) = U^{\dagger}_{L,R} 
\left( \begin{array}{c}
\psi_{L,R}\\
\chi^{u}_{L,R}
 \end{array}\right)
\label{eq:matdef}
\end{equation}
where $U_{L,R}$ are unitary matrices, so defining the 
three mass eigenstates: $\omega^1$, $\omega^2$ and 
$\chi^{d}$ with masses 

\begin{equation}
M_{\omega_1,\omega_2} = \frac{1}{2}\left[ (m_1+m_2) \mp \sqrt{(m_2-m_1)^2 + 4 c^2 v^2}\right]
\quad {\rm and } \quad 
M_{\chi} = m_1,
\label{m1m2-21}
\end{equation}
$M_{\omega_1} < M_{\chi} < M_{\omega_2}$, in most of the parameter space.

The gauge interactions with the SM fields are described by the usual
coupling with the SM fields are introduced via covariant derivatives
\begin{equation}
{\cal L_{\rm I}^{\rm 2+1}} = 
i \overline \psi \gamma^\mu \left( \partial_\mu -i g'y \, B_\mu \right) \psi + i \bar \chi \gamma^\mu 
\left( \partial_\mu - i g \,W_\mu^a 
T^a -i g'\hat{y} \, B_\mu \right) \chi , 
\label{eq:gauge-2+1}
\end{equation}
where $g'=e/c_W$, $g=e/s_W$ are the SM couplings. One can show that 
the neutral current lagrangian will be 
\begin{eqnarray}
{\cal L_{\rm NC}^{\rm 2+1}} =& e(\hat y-\frac{1}{2}) \, \bar \chi^d
\gamma_\mu \chi^d \, A^\mu + (-\hat y g' \,s_W - \frac{1}{2} \,g \,c_W)\,
\bar \chi^d \gamma_\mu \chi^d\, Z^\mu \nonumber \\ 
+ & \bar \omega \left [  U^\dagger_L \left(\begin{array}{cc}  -(\hat y + \frac{1}{2}) g'\,s_W  & 0
    \\ 0 & \frac{g}{2}c_W -\hat{y}g'\, s_W
\end{array} \right) U_L P_L + (L\to R)
\right] \gamma_\mu \omega \, Z^\mu \nonumber \\
 + & e\, (\hat y+\frac{1}{2})\, \bar \omega \gamma_\mu \omega \, A^\mu, 
\label{eq:nc-2+1}
\end{eqnarray}
and the charged current one
\begin{eqnarray}
{\cal L_{\rm CC}^{\rm 2+1}} =\frac{g}{\sqrt{2}} \bar \omega \gamma^\mu 
\left[ U_L^\dagger P_L + U_R^\dagger P_R\right] \widetilde{W}^{+T}_\mu \chi^d + \rm h.c.,
\label{eq:cc-2+1}
\end{eqnarray}
where we define $\widetilde{W}^{+}_{\mu} \equiv (0 \quad W^+_\mu)$.

\subsection{\bf Triplet-doublet model}

We will consider the following mass lagrangian for the new states
\begin{equation}
{-\cal L_{\rm mass}^{\rm 3+2}} = c \left ( \overline{\psi_R} \, \chi_L H +
\overline{\psi_L} \, \chi_R H \right)   + m_1 \, \overline \chi_L \chi_R + m_2 \, \overline \psi_L 
\psi_R + \rm h.c.   \, , 
\label{eq:hm-3+2}
\end{equation}
where $c$ is their coupling to the SM Higgs field and $m_1$ and $m_2$ their 
vector-like masses. This gives rise, after electroweak symmetry breaking,   
to the following mass matrix
\begin{equation}
M_{3+2} = (\bar \psi_R^u\, \bar \chi_R^a \,  
\bar \psi_R^d \,  \bar \chi_R^b \, 
\bar \chi_R^c) \left( \begin{array}{ccccc}
m_2 & c v & 0 & 0 & 0 \\
c v &  m_1 & 0 & 0 & 0 \\
0 & 0 & m_2 & -c\frac{v}{\sqrt{2}} & 0 \\
0 & 0 & -c\frac{v}{\sqrt{2}} & m_1 & 0 \\
0 & 0 & 0 & 0 & m_1
\end{array}
\right) 
\left( \begin{array}{c}
\psi_L^u \\
\chi_L^a\\
\psi_L^d \\
\chi_L^b \\
\chi_L^c 
 \end{array}\right)\, ,
\label{eq:mass-3+2}
\end{equation}
 where  the doublet and triplet  read
\begin{equation}
\psi = \begin{pmatrix}
\psi^u \cr \psi^d
\end{pmatrix} \, ,
\quad \quad
\chi = \begin{pmatrix}
{\chi^b \over \sqrt{2}} & \chi^a \cr \chi^c & {\chi^b \over \sqrt{2}}
\end{pmatrix}\, .
\end{equation}

To diagonalize $M_{3+2}$ we introduce the following transformations  

\begin{equation}
\omega_{L,R} \equiv
\left( \begin{array}{c}
\omega^{1}_{L,R}\\
\omega^{2}_{L,R}
 \end{array}\right) = U^{\dagger}_{L,R} 
\left( \begin{array}{c}
\psi^{u}_{L,R}\\
\chi^{a}_{L,R}
 \end{array}\right)
\quad \quad
\xi_{L,R}\equiv
\left( \begin{array}{c}
\xi^{1}_{L,R}\\
\xi^{2}_{L,R}
 \end{array}\right) = V^{\dagger}_{L,R} 
\left( \begin{array}{c}
\psi^{d}_{L,R}\\
\chi^{b}_{L,R}
 \end{array}\right)
\label{eq:rots}
\end{equation}
where $U_{L,R},V_{L,R}$ are unitary matrices, so defining the 
five mass eigenstates: $\omega^1$, $\omega^2$, $\xi^1$, $\xi^2$ and 
$\chi=\chi^{c}$ with masses 

\begin{equation}
M_{\omega_1,\omega_2} = \frac{1}{2} \left[ (m_1+m_2) \mp \sqrt{(m_2-m_1)^2 + 4 c^2 v^2}
\right] \nonumber
\end{equation}
\begin{equation}
M_{\xi_1,\xi_2} = \frac{1}{2}\left[ (m_1+m_2) \mp \sqrt{(m_2-m_1)^2 + 2 c^2 v^2}\right] \nonumber
\end{equation}
\begin{equation}
M_{\chi} = m_1 
\label{m1m2-32}
\end{equation}
$M_{\omega_1} < M_{\xi_1} < M_{\chi}< M_{\omega_2} < M_{\xi_2}$, in
most of the parameter space.

The gauge interactions with the SM fields are described again in the 
usual way by 
\begin{equation}
{\cal L_{\rm I}^{\rm 3+2}} = 
i \overline \psi \gamma^\mu \left( \partial_\mu - i g \, W_\mu^a  
T^a -i g'y \, B_\mu \right) \psi + i \bar \chi \gamma^\mu 
\left( \partial_\mu - i g W_\mu^a 
T^a -i g'\hat{y} \, B_\mu \right) \chi , 
\label{eq:gauge-3+2}
\end{equation}
giving rise to the neutral current lagrangian 
\begin{eqnarray}
{\cal L_{\rm NC}^{\rm 3+2}}=& e(\hat y-1) \, \bar \chi \gamma_\mu \chi \, A^\mu + 
(-g' \, \hat{y} \,s_W - g \,c_W)\, \bar \chi \gamma_\mu \chi\, Z^\mu \nonumber \\
 + & \bar \omega \left [ 
U^\dagger_L \left(\begin{array}{cc}
\frac{g}{2}\, c_W - yg'\, s_W & 0 \\
0 & g\, c_W- \hat{y}g'\, s_W  
\end{array} \right) U_L P_L + (L\to R)
\right] \gamma_\mu \omega \, Z^\mu \nonumber \\
 + & (\hat y+1) \, e \, \bar\omega \gamma_\mu \, \omega A^\mu \nonumber \\
 + & \bar \xi \left [ 
V^\dagger_L \left(\begin{array}{cc}
-\frac{g}{2}\, c_W- yg'\, s_W & 0 \\
0 & -\hat{y}\,g'\, s_W  
\end{array} \right) V_L P_L + (L\to R)
\right] \gamma_\mu \xi \, Z^\mu \nonumber \\
 + & \hat y\,e \, \bar \xi \gamma_\mu \xi \, A^\mu,
\label{eq:nc-3+2}
\end{eqnarray}
and to the charged current lagrangian

\begin{eqnarray}
{\cal L_{\rm CC}^{\rm 3+2}}=& 
 g \left (\bar \omega \; \bar \xi \; \bar \chi \right)\gamma^\mu
\left [
\left(\begin{array}{ccc}
0_{2\times 2} & W^+_\mu \, U_L^\dagger V'_L & 0_{2\times 1} \\
W^-_\mu \, V_L'^{\dagger}U_L & 0_{2\times 2} & V_L^{\dagger}  \widetilde{W}^{+T}_{\mu} \\
0_{1\times 2} & \widetilde{W}^{-}_{\mu} \, V_L & 0
\end{array} \right) P_L 
+ (L\to R) \right]
\left(
\begin{array}{c}
\omega \\
\xi \\
\chi
\end{array}
\right),
\label{eq:cc-3+2}
\end{eqnarray}
where $0_{n\times m}$ is a $n\times m$ zero matrix, 
$\widetilde{W}^-_{\mu} \equiv (0  \quad W^-_\mu)$, $\widetilde{W}^{+}_{\mu} \equiv (0 \quad W^+_\mu)$ and 

\begin{equation}
V_L'= \frac{1}{\sqrt{2}}\left(\begin{array}{cc}
V_{11L} & V_{12L} \\
\sqrt{2}\, V_{21L} & \sqrt{2}\, V_{22L}
\end{array}
\right).
\end{equation}

\section{$H\to \gamma \gamma$ Width}
\label{sec:hgg}

We have studied the Higgs to diphoton width in the doublet-singlet model 
and in the triplet-doublet model. In both models the ratio 
\begin{equation}
R_{\gamma\gamma} = \frac{\Gamma(H \to \gamma\gamma)}{\Gamma^{\rm SM}(H\to \gamma\gamma)}\, ,
\label{eq:rgammagamma}
\end{equation}
between the width of $H\to \gamma \gamma$ with extra states and the
width of $H\to \gamma \gamma$ in the SM have the feature that fixing 
the lightest new fermion mass the largest enhancement will be achieved for  
$m_1=m_2$. This is illustrated in Fig.~\ref{fig:whym1m2}.
These plots also show the symmetry between $m_1$ and $m_2$ that can 
be also seen from Eqs.~(\ref{m1m2-21}) and (\ref{m1m2-32}). 

In Fig.~\ref{fig:HAA21} we show the allowed regions at 68\%, 95\% and
99\% CL, according to Ref.~\cite{Corbett:2012dm}, for the ratio
$R_{\gamma \gamma}$ in the plane $(m_1=m_2) \times c$ for the
doublet-singlet model and for several $\omega$ charges.  We also show
some iso-lines corresponding to the mass of the lightest new particle,
either $\omega_1$ or $\omega_2$, depending on the parameter
values. We will simply call this lightest mass $M_{\rm light}$.  The
region where $M_{\rm light} \lsim$ 100 GeV is already excluded by LEP
data~\cite{Achard:2001qw}. As a reference we also display some
iso-lines for fixed values of $R_{\gamma \gamma}$.

In the doublet-singlet model there is only one electric charge at play
in the extra states contribution, so there is no sensitivity to the
sign of the charge.  However, as $m_1, m_2$ and $c$ vary
one can have a complete cancellation of their contribution to 
$H\to \gamma \gamma$, and for a small parameter region, even have a 
smaller $\Gamma(H \to \gamma \gamma)$ that the SM one. For this reason, 
we have two disconnected regions that are consistent with  
the allowed range of $R_{\gamma\gamma}$. 
On the top panel of Fig.~\ref{fig:HAA21} those two regions are very close 
together and for the most part concentrated at $M_{\rm light} \lsim 100$ GeV.
As we increase the absolute charge of $\omega$ we start to see these regions 
being pulled apart and away from $M_{\rm light} \lsim 100$ GeV (see middle 
panel of Fig.~\ref{fig:HAA21}) until one of the regions disappear from 
the plot.
 
For a coupling to the Higgs of ${\cal O}(1)$, hypercharge $y=1$ and $M_{\rm light}> 100$ 
GeV there is compatibility only if the future LHC data show a decrease of 
$\Gamma(H \to \gamma \gamma)$ to a value at most 25\% higher than the SM one, 
{\it i.e.}, in the 99\% CL allowed region by the current LHC data.
For $y>1$ there are solutions even inside the 68\% CL, for $M_{\rm light}$ that 
can be as high as a few hundred of GeV, well inside the LHC discovery reach.

In Fig.~\ref{fig:HAA32} we show the allowed regions at 68\%, 95\% and
99\% CL, according to Ref.~\cite{Corbett:2012dm}, for the ratio
$R_{\gamma \gamma}$ in the plane $(m_1=m_2) \times c$ for the
triplet-doublet model and for several $\omega$ charges.  We also show
some iso-lines corresponding to the mass for the lightest new
particle, eitheir $\omega_1$ or $\xi_1$, depending on the parameter
values. Again we will call this lightest mass $M_{\rm light}$.  The
region where $M_{\rm light} \lsim$ 100 GeV is already excluded by LEP
data~\cite{Achard:2001qw}. As a reference we also display some
iso-lines for fixed $R_{\gamma \gamma}$.

In the triplet-doublet model there are four states and two 
different electric charges at play, so different charge combinations 
will give rise to a more interesting behavior. As before, and for 
the same reasons explained above,  we have two disconnected regions that are 
consistent with  the allowed $R_{\gamma\gamma}$. 
However, as we increase the charges these regions swap places.

In Fig.~\ref{fig:HAA32}, the charges of $(\omega,\xi)$ are,
respectively, from left to right and top to bottom, $(-3,-4),
(-2,-3), (-1,-2), (0, -1), (2,1)$ and $(3,2)$.  For large negative
charges we can see only one of those regions (top panels).  As we
increase the charges, we start to see both regions, although most of
the second allowed region is forbidden by the LEP limit (middle
panels).  Finally the two regions crossover and swap places (bottom
panels).  We do not show here the case $y=1/2$ because it is almost
exactly the same as the top panel of Fig.~\ref{fig:HAA21}, since
$\omega$ have charge 1 and $\xi$ charge 0 in this case.

For a coupling to the Higgs of ${\cal O}(1)$, the compatibility 
region where there is a solution for 
$R_{\gamma\gamma}$ at 68\% and 95\% CL strongly depends on the $y$ 
value. Again the lightest particle can be within the LHC 
discovery range.

\section{Oblique Parameters S and T}
\label{sec:STU}

We now examine the contributions of these new states to the oblique
parameters. In Fig.~\ref{fig:ST21} we show the allowed regions at
68\%, 95\% and 99\% CL for $S$ (left panel) and $T$ (right panel) for
the doublet-singlet model with $y=1$ to illustrate the $S$ and $T$
separate behavior.  We do not show a plot for $U$, since it poses
practically no additional bound on the parameter space.  $S$ has two
preferred regions, reflecting the same degeneracy we have seen before
for $R_{\gamma \gamma}$. $T$, for $c \lsim$ 0.5 allows for any values
of $m_1=m_2$, but as $c$ increases lower values of $m_1=m_2$ become
forbidden. This is because $T$ is sensitive to the mass splitting of
the doublet and prefers thus smaller mixing.

In Fig.~\ref{fig:ST32} we show the the allowed regions at 68\%, 95\%
and 99\% CL for $S$ (left panel) and $T$ (right panel) for the
triplet-doublet model for $y=-5/2$ to illustrate the $S$ and $T$
separate behavior. Again, we do not show a separate plot for $U$.
Here the second $S$ region becomes a thin strip, while for $T$ the
behavior is very similar to the doublet-singlet case.

Now we will combine $S$ and $T$ using the $\chi^2$ function described
in Eq.~(\ref{eq:chi2}) of Sec.~\ref{sec:model} allowing for
correlations.  In Fig.~\ref{fig:ST21-comb} we show the combined region
allowed for the fit of $S$ and $T$ for $y=1$ and $m_1=m_2$ (left
panel) and for $c=y=1$ (right panel) for the doublet-singlet
case. When we compare this with with Fig.~\ref{fig:ST21} we see that
the $T$ parameter pulls the combined allowed region down. The plot on
the right also shows the correlation between $m_1$ and $m_2$ allowed
by $T$.  In Fig.~\ref{fig:ST32-comb} we show the combined region
allowed for the fit of $S$ and $T$ for $y=-5/2$ and $m_1=m_2$ (left
panel) and for $c=1$ and $y=-5/2$ (right panel) for the
triplet-doublet case.

In Fig.~\ref{fig:TOT-2+1} we see that for the case where $y=1$
there is great tension between the region that is preferred by 
$R_{\gamma \gamma}$ and the region allowed by $S$ and $T$ parameters.
This is expected  because $T$ prefers a region where there is a small 
breaking of SU(2)$_L$ while $R_{\gamma \gamma}$ is mostly enhanced for 
large mixing. The same general behavior can be observed in 
Fig.~\ref{fig:TOT-3+2} for $y=-5/2$ in the triplet-doublet model, 
although in this case, the tension is alleviated.

\section{Production at the LHC}
\label{sec:production}
These new charged states could be produced at the LHC and we have calculated 
the cross section for pair production of the lightest fermion
in the doublet-singlet model, and correspondingly in the triplet-doublet model.

This was done for p-p collisions at the LHC running at a center of
mass energy of 8 TeV, using CalcHEP~\cite{calchep} with
CTEQ6L~\cite{Pumplin:2002vw} parton distribution functions for the
proton and imposing the following loose cuts: $-2.1 < \eta < 2.1$ and $p_T>
40$ GeV for each fermion.

For both models, the main contributions to pair production of the
lightest particle come from neutral currents $Z$ and $\gamma$ exchange
in the $s$ channel.

In Fig.~\ref{fig:crosspair} the results of our calculations for the
production cross section as a function of the mass of the lightest
particle, for $y=1$ for the doublet-singlet model and $y=-3/2$ for the
doublet-triplet model.  We also include in Fig.~\ref{fig:crosspair},
the current bounds on the cross-section on stau from CMS
\cite{stauspairproduction}.

These stringent bounds in the lower mass region could lead to some
indication  of the possible charges of these new fermions.  For example,
for masses $> 300 $ GeV, together with the electroweak precision tests,
Figs.~\ref{fig:HAA21} and \ref{fig:HAA32} point to 
specific charges that are preferred for the current allowed regions of
$R_{\gamma \gamma}$. Additional structure to this sector, on the other
hand, could permit one to evade them.

\begin{figure*}[!t]
\begin{center}
 \includegraphics[width=0.49\textwidth]{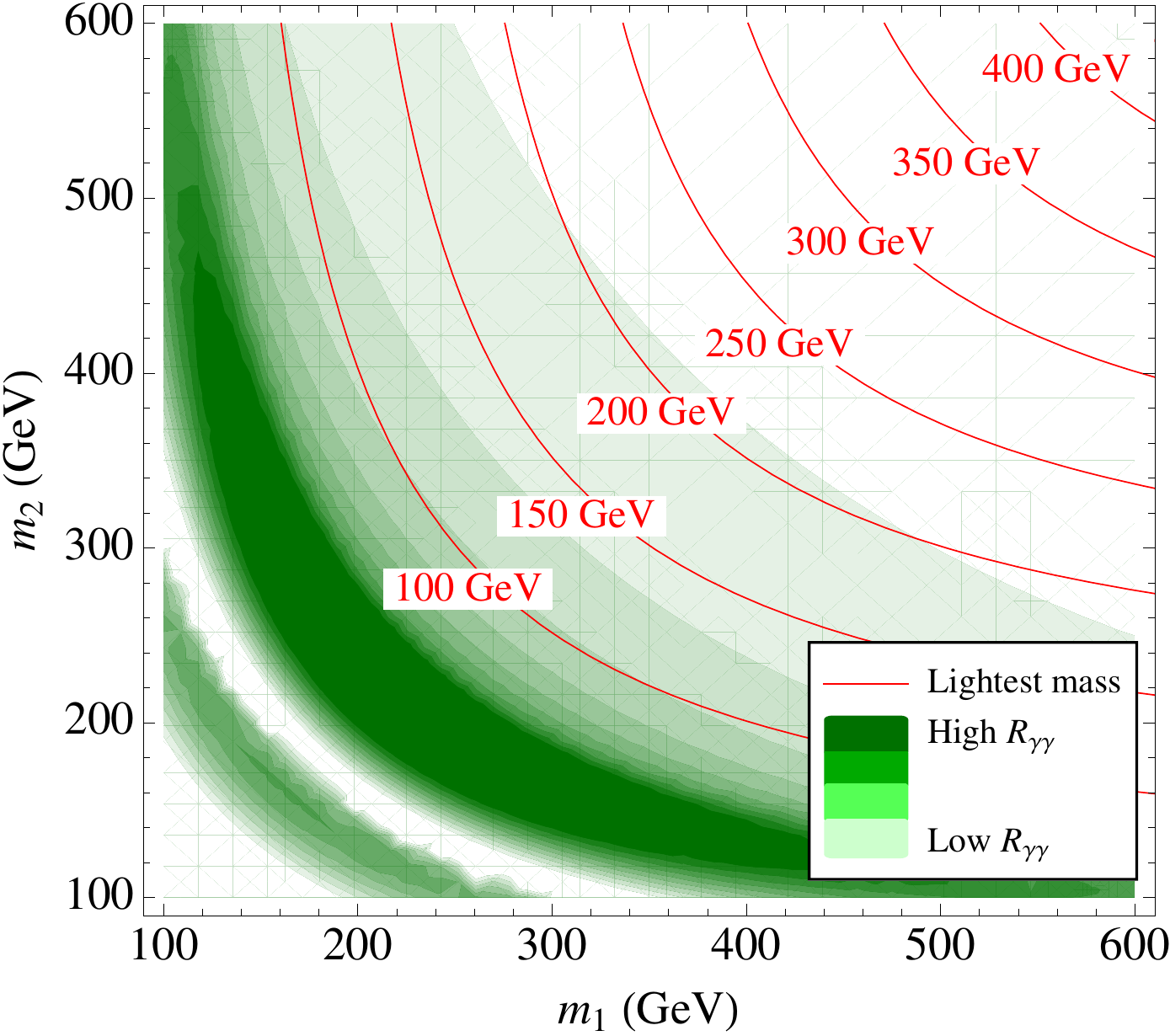}
 \includegraphics[width=0.49\textwidth]{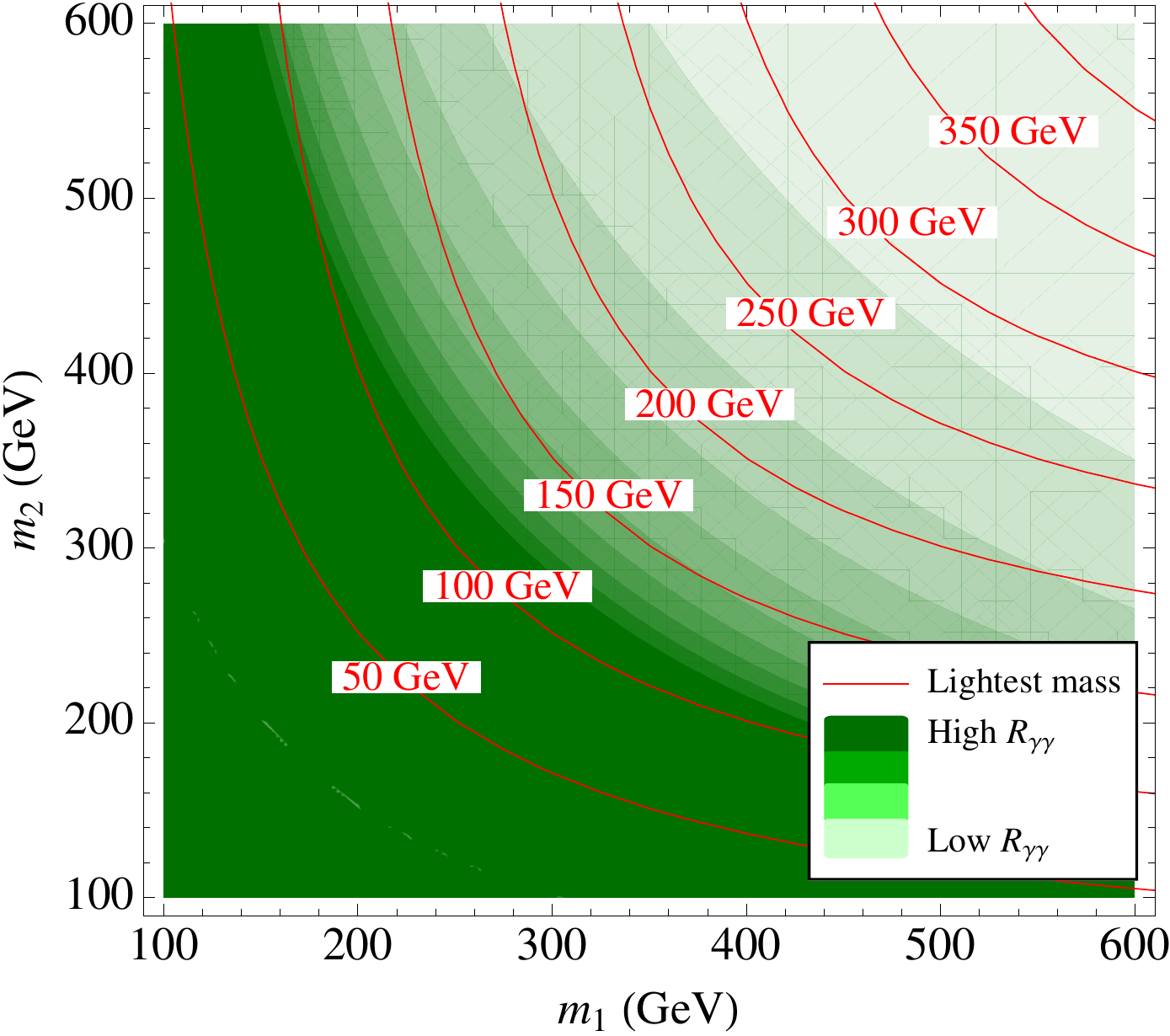}
\vspace{-2mm}
\end{center}
\vspace{-0.1cm}
\caption{Iso-contours  of the ratio $R_{\gamma \gamma}$ in the plane 
$m_1$ versus $m_2$ for $c=1$. We also show in the same plot the iso-lines that 
correspond to a fixed value of the lightest new charged fermion mass.
The doublet-singlet model behavior with $y=1$ is illustrated on the left  and 
the triplet-doublet model with $y=-5/2$ on the right panel.}
\label{fig:whym1m2}
\end{figure*}

\begin{figure*}[!t]
\begin{center}
 \includegraphics[width=0.45\textwidth]{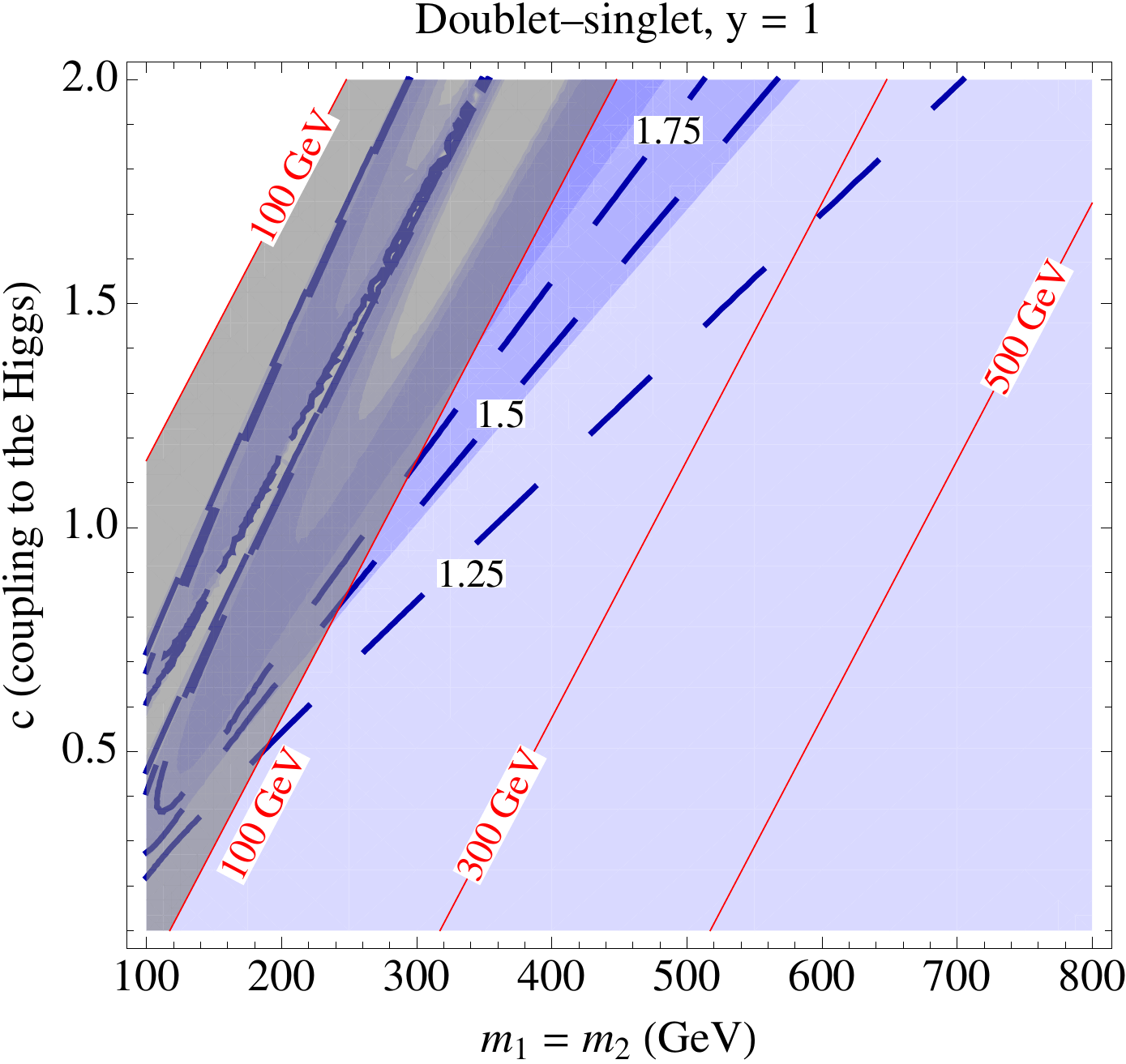}\\
 \includegraphics[width=0.45\textwidth]{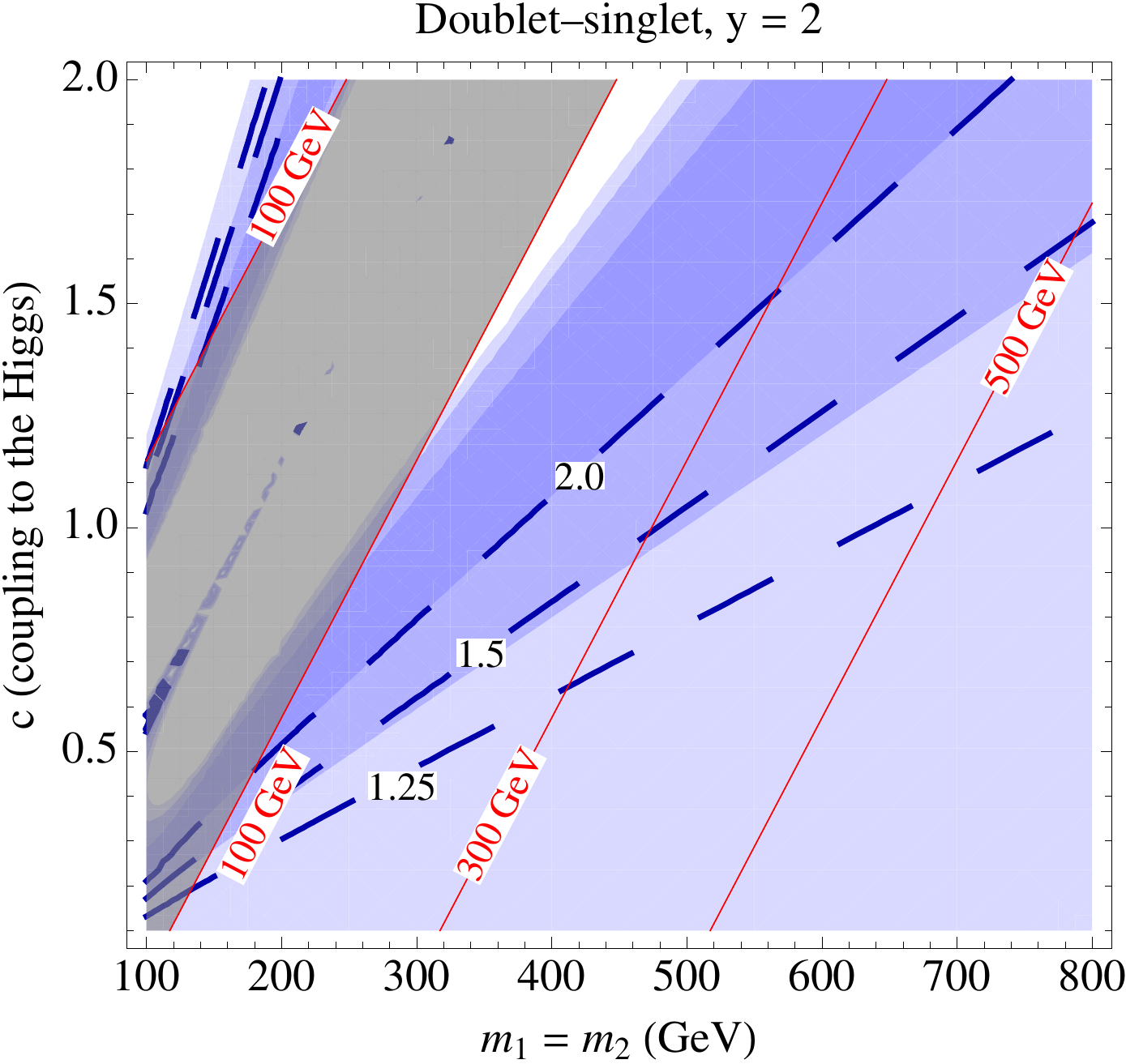}\\
 \includegraphics[width=0.45\textwidth]{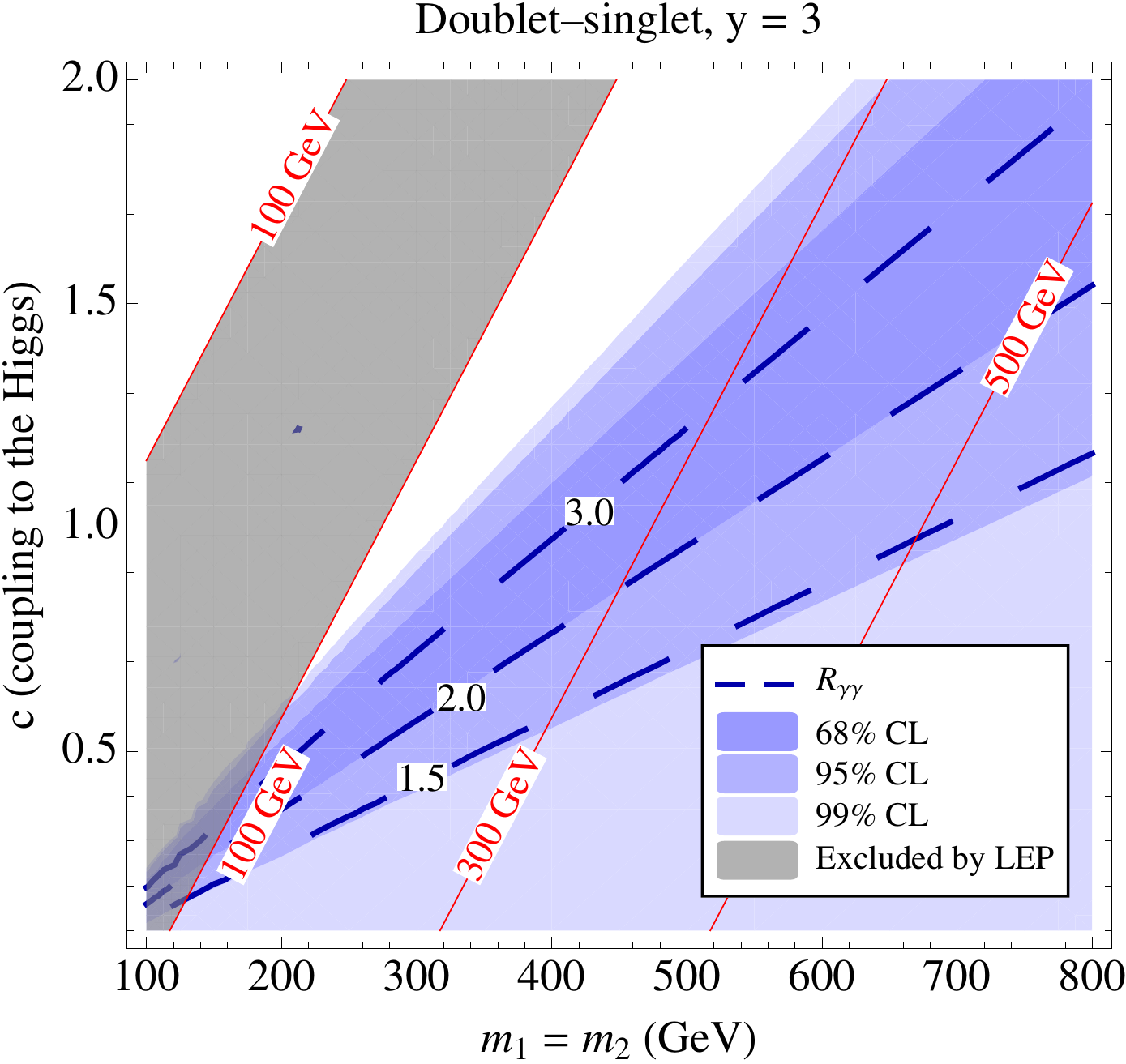}
\vspace{-2mm}
\end{center}
\vspace{-0.1cm}
\caption{Regions in the plane $(m_1=m_2) \times c$ allowed by the fit
  to the combined Higgs data for $R_{\gamma \gamma}$, according to
  Ref.~\cite{Corbett:2012dm} at 68\% (darker blue), 95\% (intermediate
  blue) and 99\% (lighter blue) CL, for the doublet-singlet model.  We
  also show some iso-lines of constant mass for the lightest new
  particle, $M_{\rm light}$, depending on the parameter values. The
  region excluded by LEP, for $M_{\rm light} \lsim$ 100 GeV, is shown
  in grey. As a reference we also show some isolines that correspond
  to $R_{\gamma \gamma}=1.25, 1.5, 1.75, 2.0$ and $3.0$.  On the top,
  middle and lower panels we show the cases $y=1,2$ and $3$, which
  corresponds to $\omega$ with charge $\pm 1$, $\pm 2$ and $\pm 3$,
  respectively. }
\label{fig:HAA21}
\end{figure*}

\begin{figure*}[!t]
\begin{center}
 \includegraphics[width=0.4\textwidth]{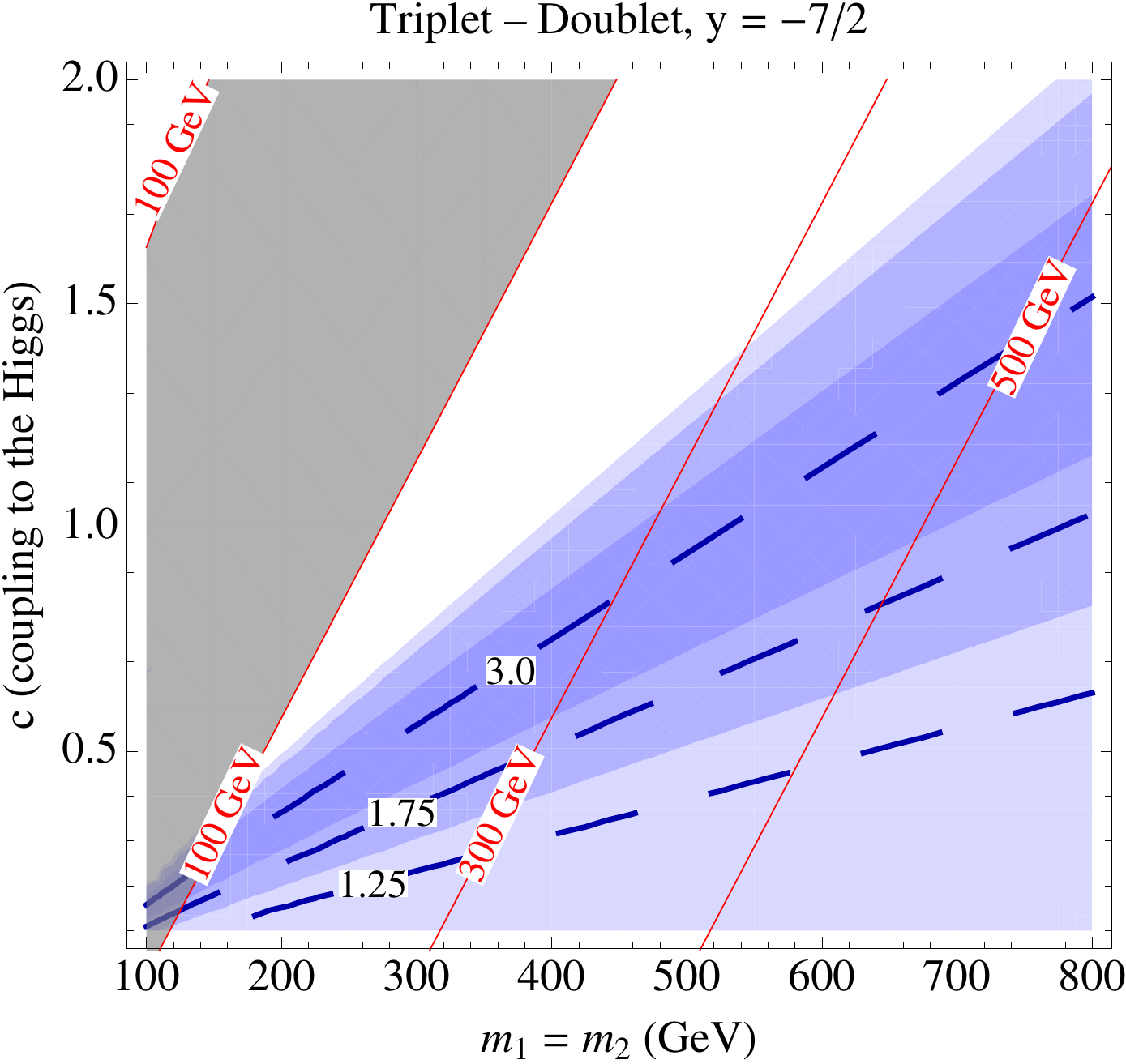}
 \includegraphics[width=0.4\textwidth]{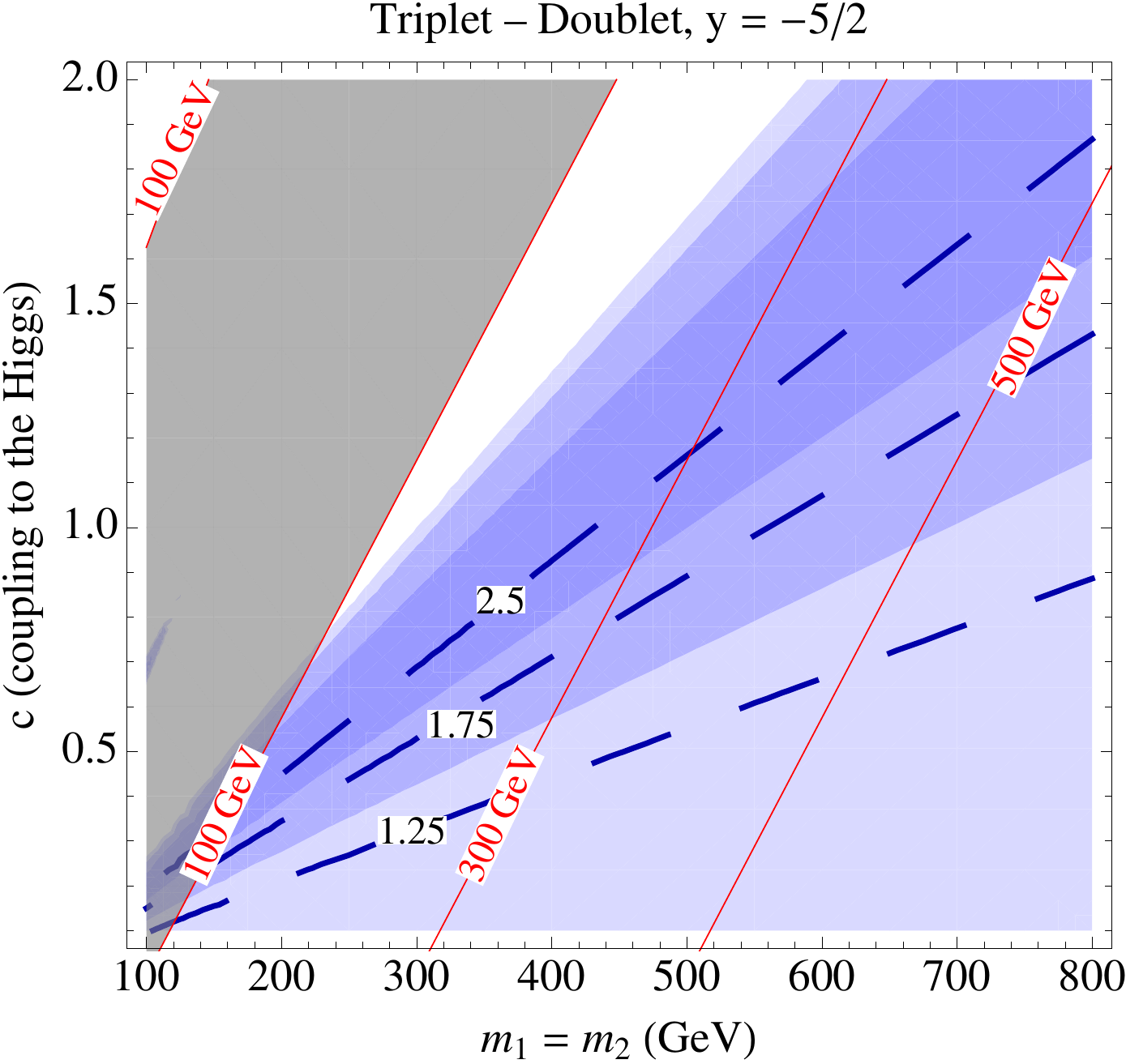}\\
 \includegraphics[width=0.4\textwidth]{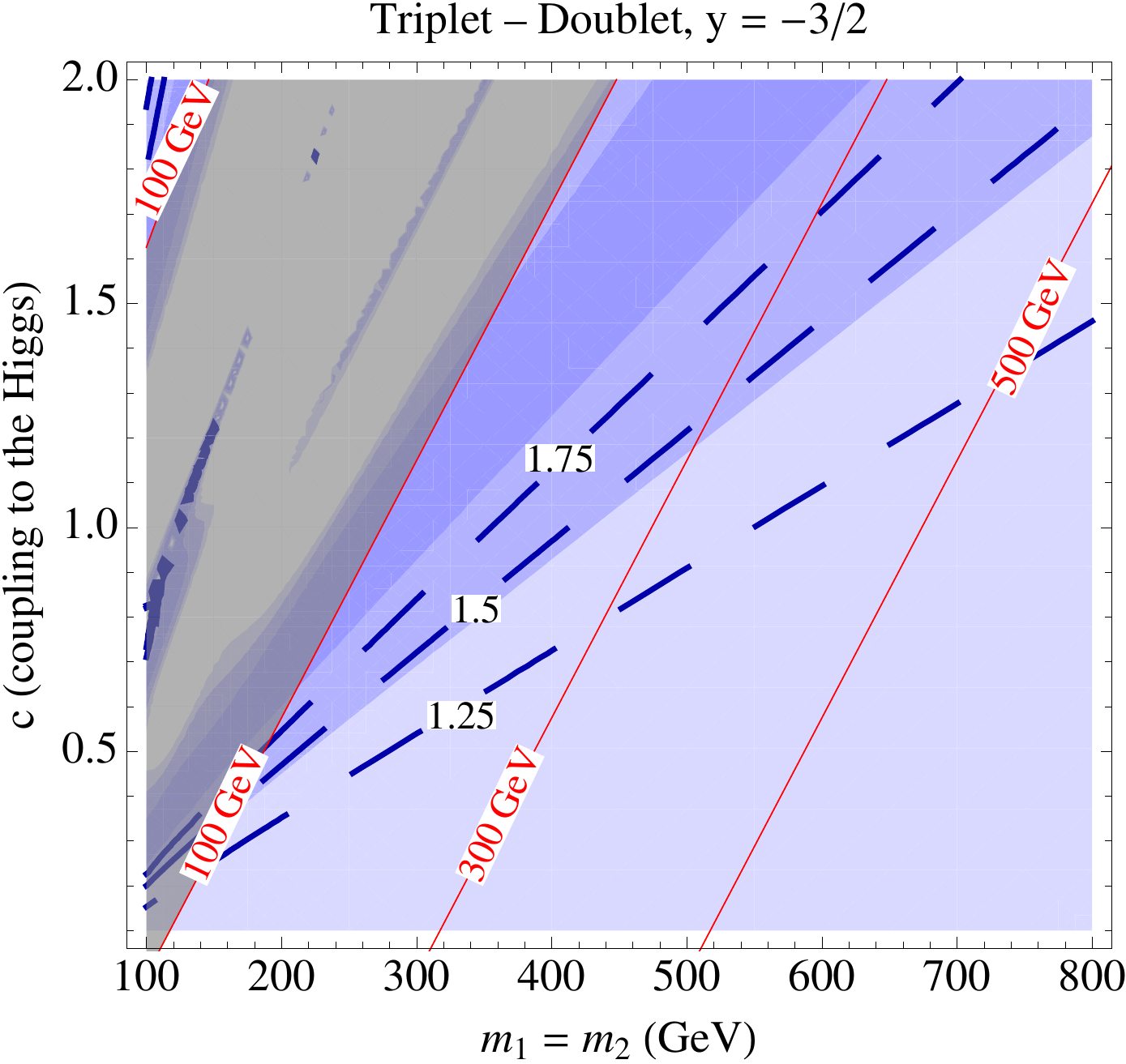}
 \includegraphics[width=0.4\textwidth]{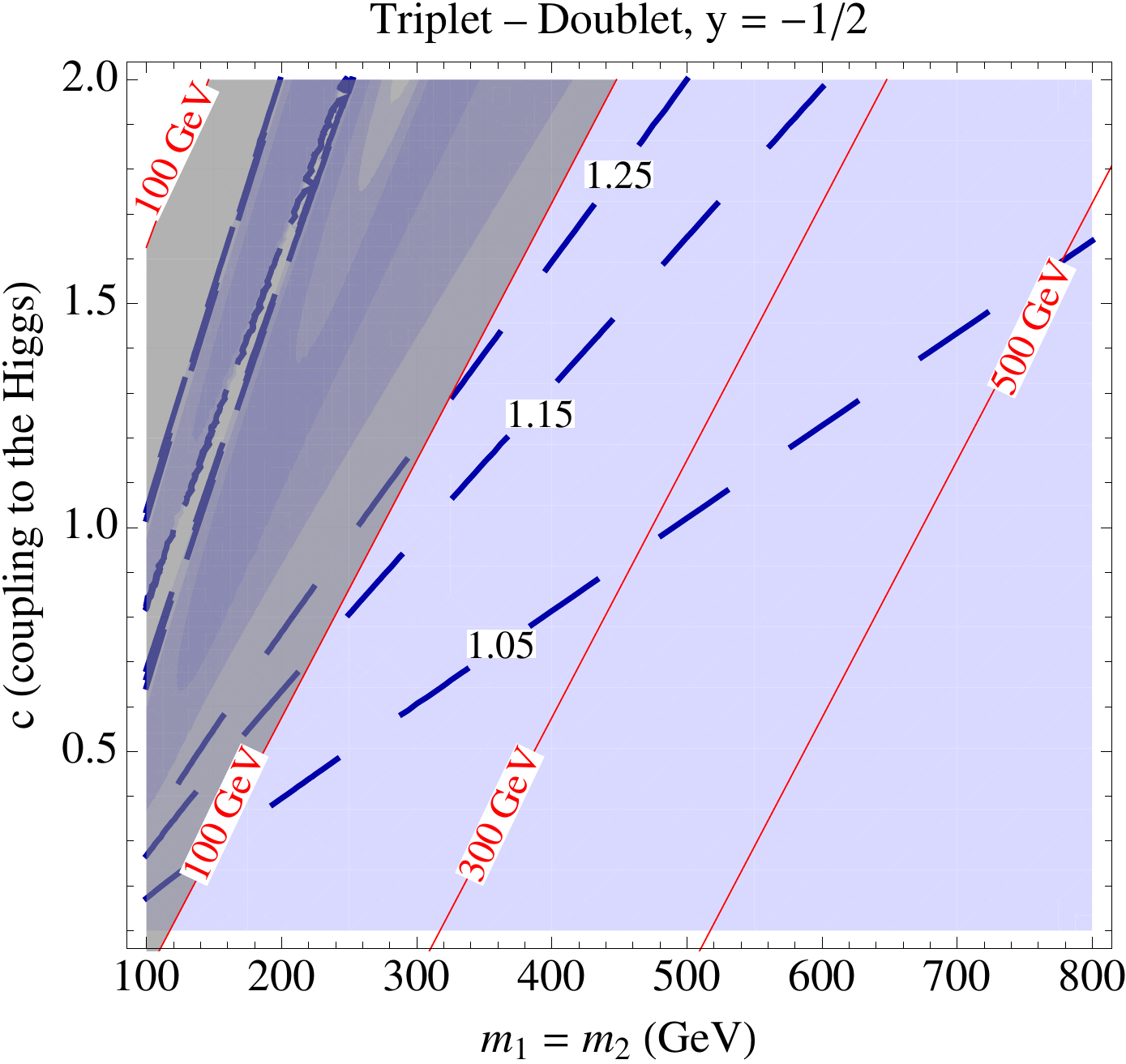}\\
\includegraphics[width=0.4\textwidth]{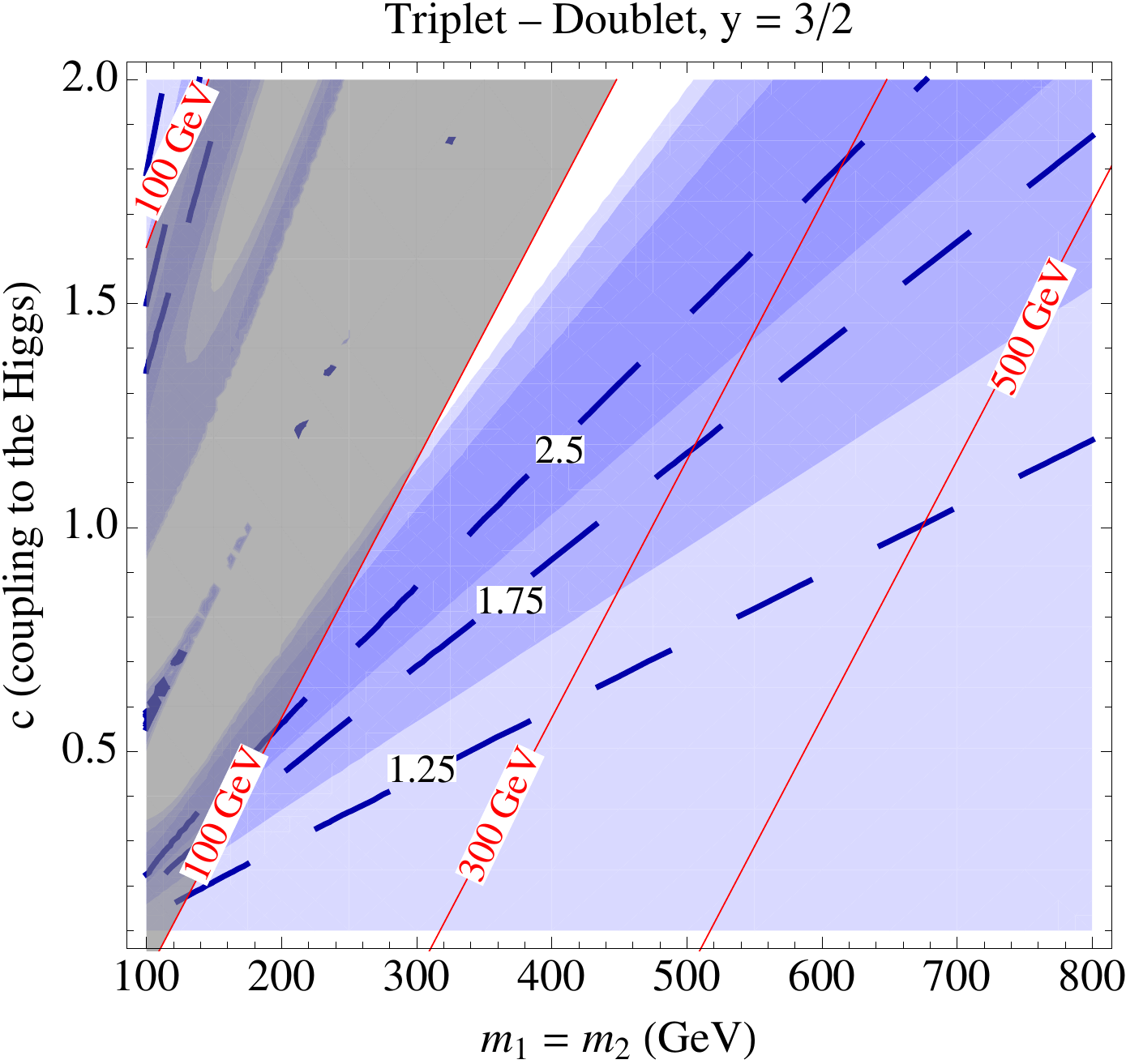}
\includegraphics[width=0.4\textwidth]{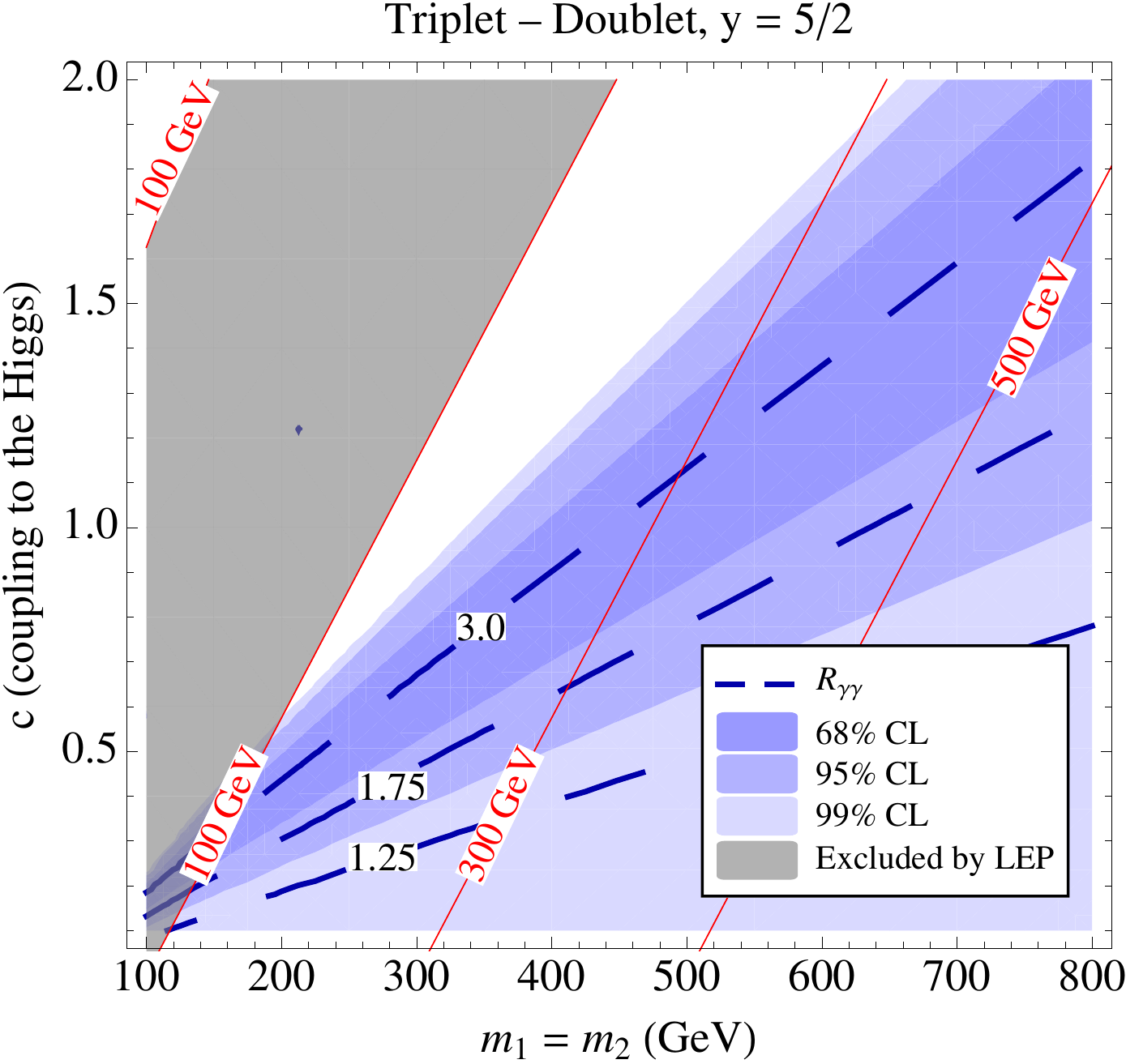}
\vspace{-2mm}
\end{center}
\vspace{-0.1cm}
\caption{Regions in the plane $(m_1=m_2) \times c$ allowed by the fit to
  the combined Higgs data for $R_{\gamma \gamma}$, according to
  Ref.~\cite{Corbett:2012dm} at 68\% (darker blue), 95\% (intermediate
  blue) and 99\% (lighter blue) CL, for the triplet-doublet model.  We
  also show some iso-lines of constant mass for the lightest new
  particle, $M_{\rm light}$, depending on the
  parameter values. The region excluded by LEP, for $M_{\rm light}$
  $\lsim$ 100 GeV, is shown in grey. As a reference
  we also show some isolines that correspond to $R_{\gamma
    \gamma}=1.05, 1.15, 1.25, 1.5, 1.75$ and $2.5$.  On the top panel
  we show on the left (right) the case $y=-7/2$ ($y=-5/2$) that
  correspond to $\omega$ and $\xi$ with charges $-3$ and $-4$ ($-2$
  and $-3$), respectively. On the middle panel
  we show on the left (right) the case $y=-3/2$ ($y=-1/2$) that
  correspond to $\omega$ and $\xi$ with charges $-1$ and $-2$ ($0$
  and $-1$), respectively. On the lower panel 
  we show on the left (right) the case $y=3/2$ ($y=5/2$) that
  correspond to $\omega$ and $\xi$ with charges $2$ and $1$ ($3$
  and $2$), respectively.}
\label{fig:HAA32}
\end{figure*}

\begin{figure*}[!t]
\begin{center}
 \includegraphics[width=0.49\textwidth]{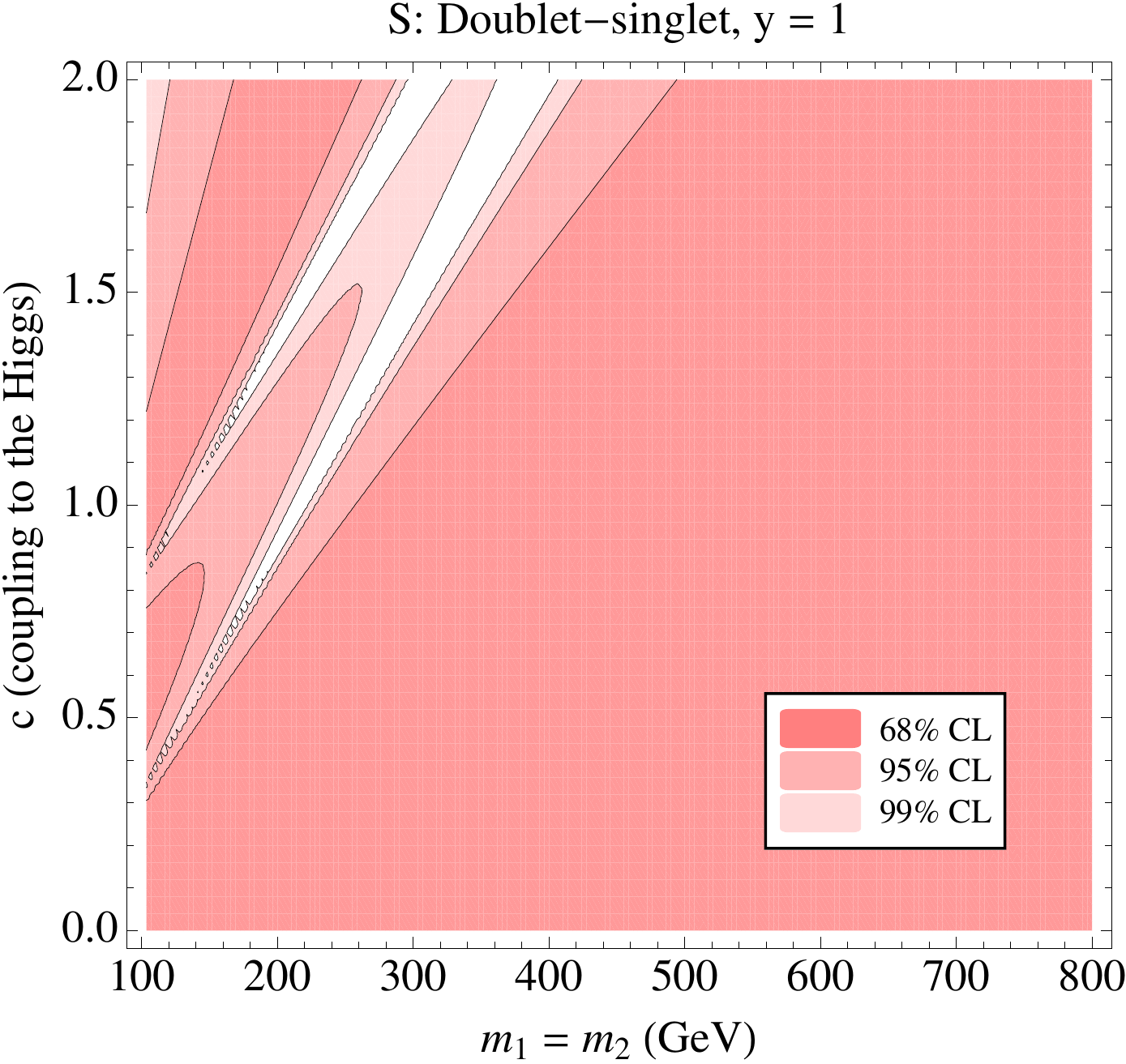}
 \includegraphics[width=0.49\textwidth]{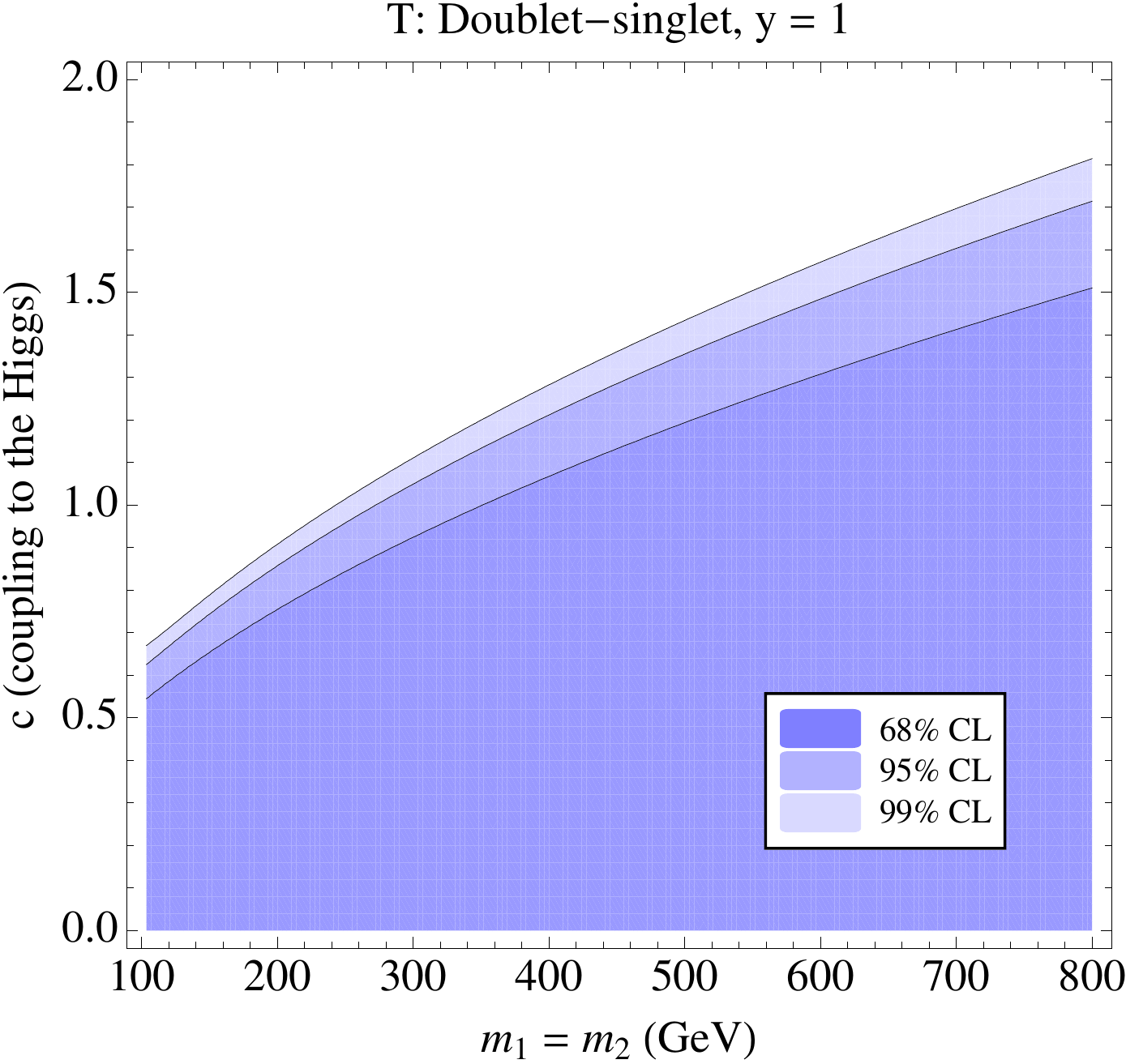}
\vspace{-2mm}
\end{center}
\vspace{-0.1cm}
\caption{Allowed regions for S (left panel) and T (right panel)
in the plane $m_1=m_2$ versus $c$, the coupling to the Higgs 
for the doublet-singlet model with $y=1$.}
\label{fig:ST21}
\end{figure*}

\begin{figure*}[!t]
\begin{center}
 \includegraphics[width=0.49\textwidth]{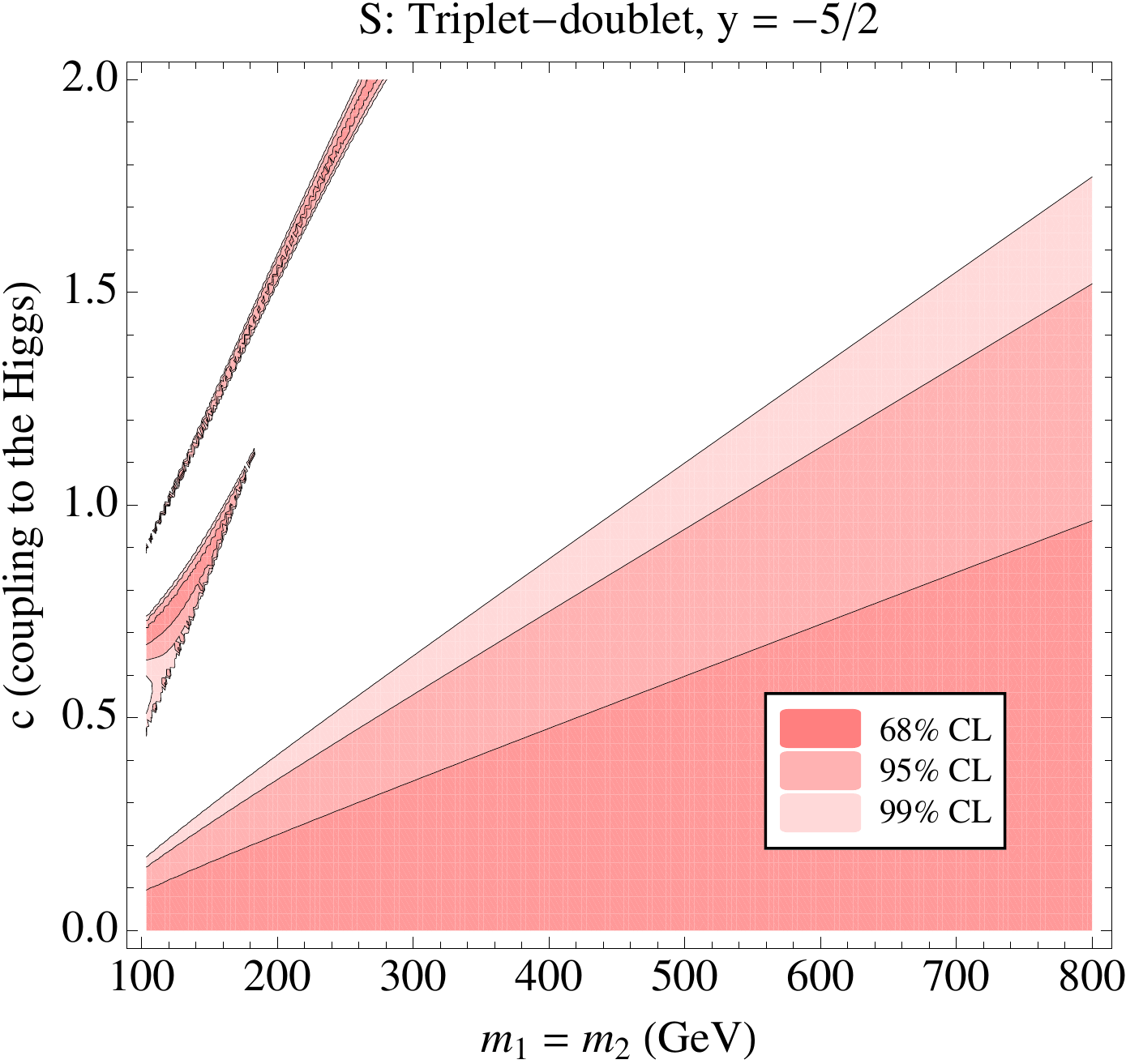}
 \includegraphics[width=0.49\textwidth]{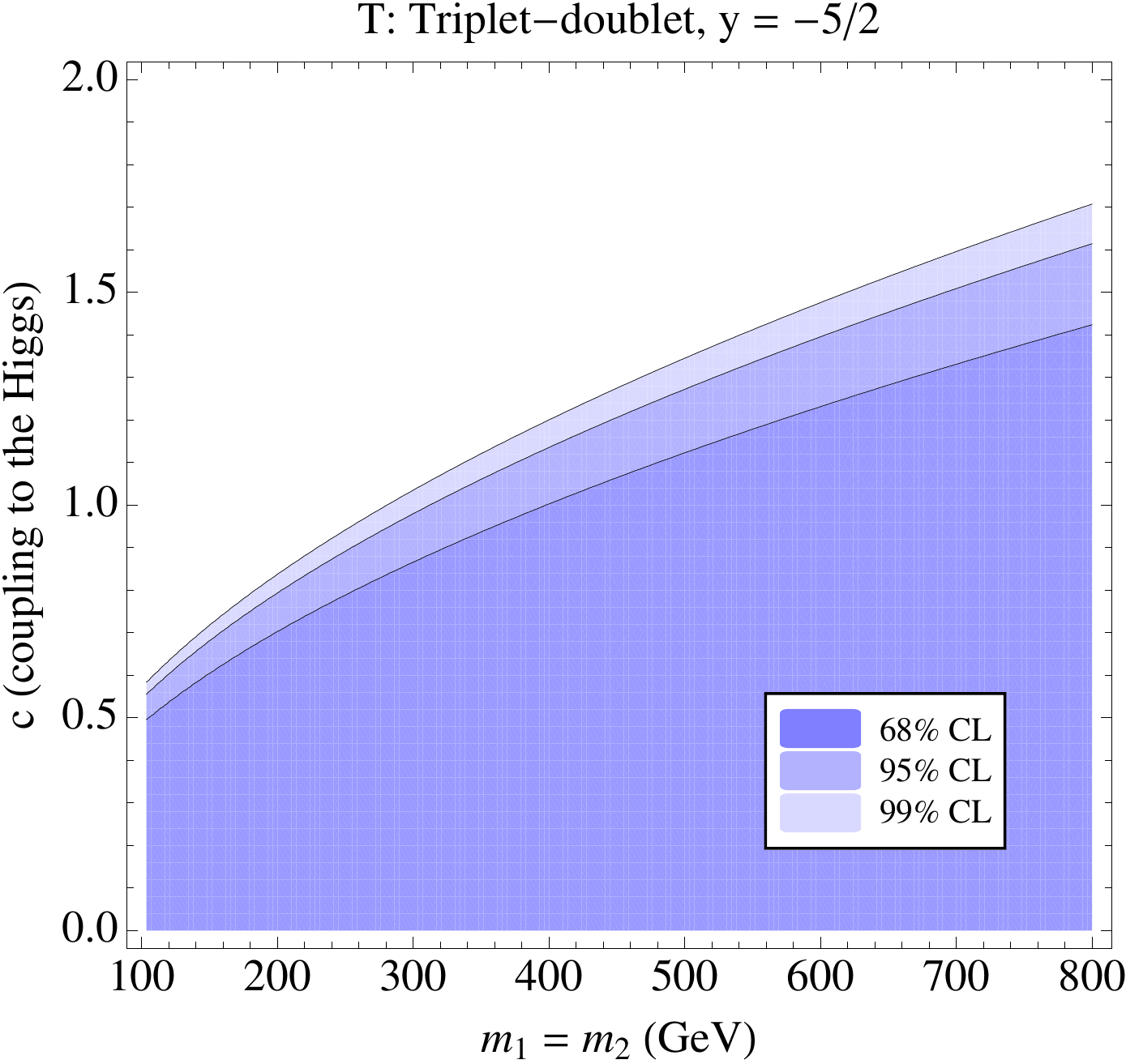}
\vspace{-2mm}
\end{center}
\vspace{-0.1cm}
\caption{Allowed regions for S (left panel) and T (right panel)
in the plane $m_1=m_2$ versus $c$, the coupling to the Higgs 
for the triplet-doublet model with $y=-5/2$.}
\label{fig:ST32}
\end{figure*}

\begin{figure*}[!t]
\begin{center}
 \includegraphics[width=0.49\textwidth]{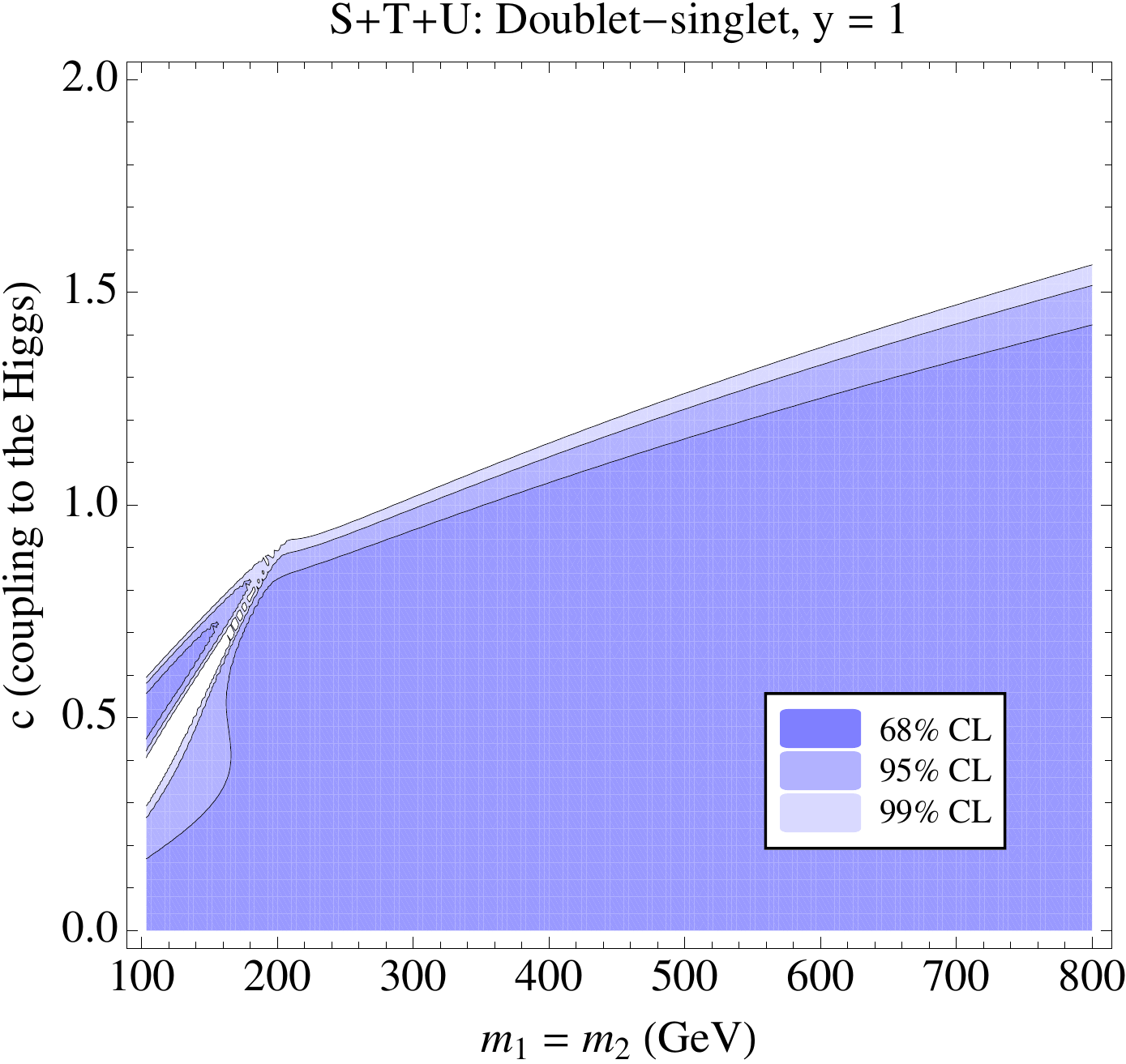}
 \includegraphics[width=0.49\textwidth]{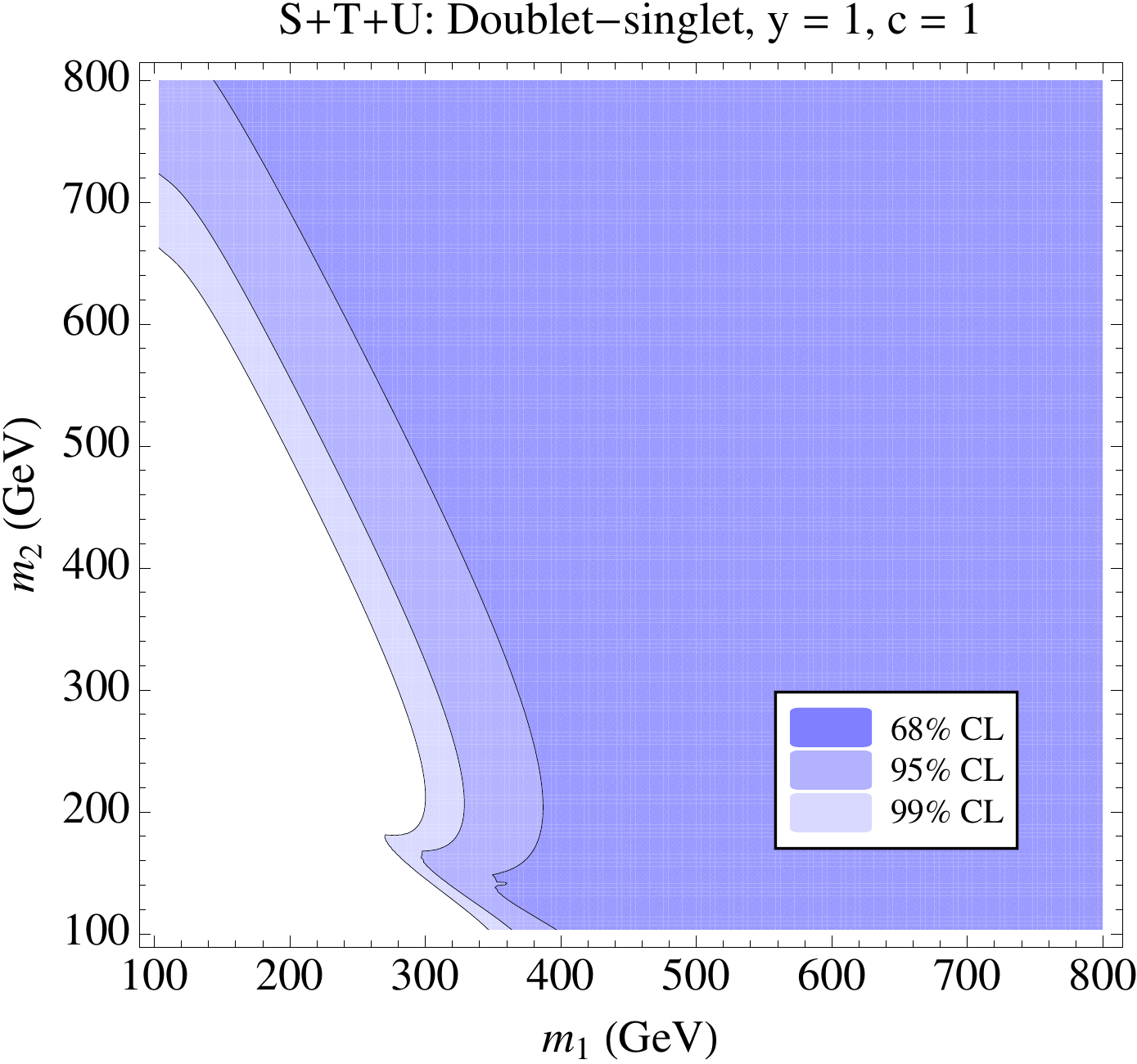}
\vspace{-2mm}
\end{center}
\vspace{-0.1cm}
\caption{Allowed region for the combined fit of S, T and U 
for $y=1$ and $m_1=m_2$ (left panel) and 
for $y=1$ and $c=1$ (right panel) in the doublet-singlet model.}
\label{fig:ST21-comb}
\end{figure*}

\begin{figure*}[!t]
\begin{center}
 \includegraphics[width=0.49\textwidth]{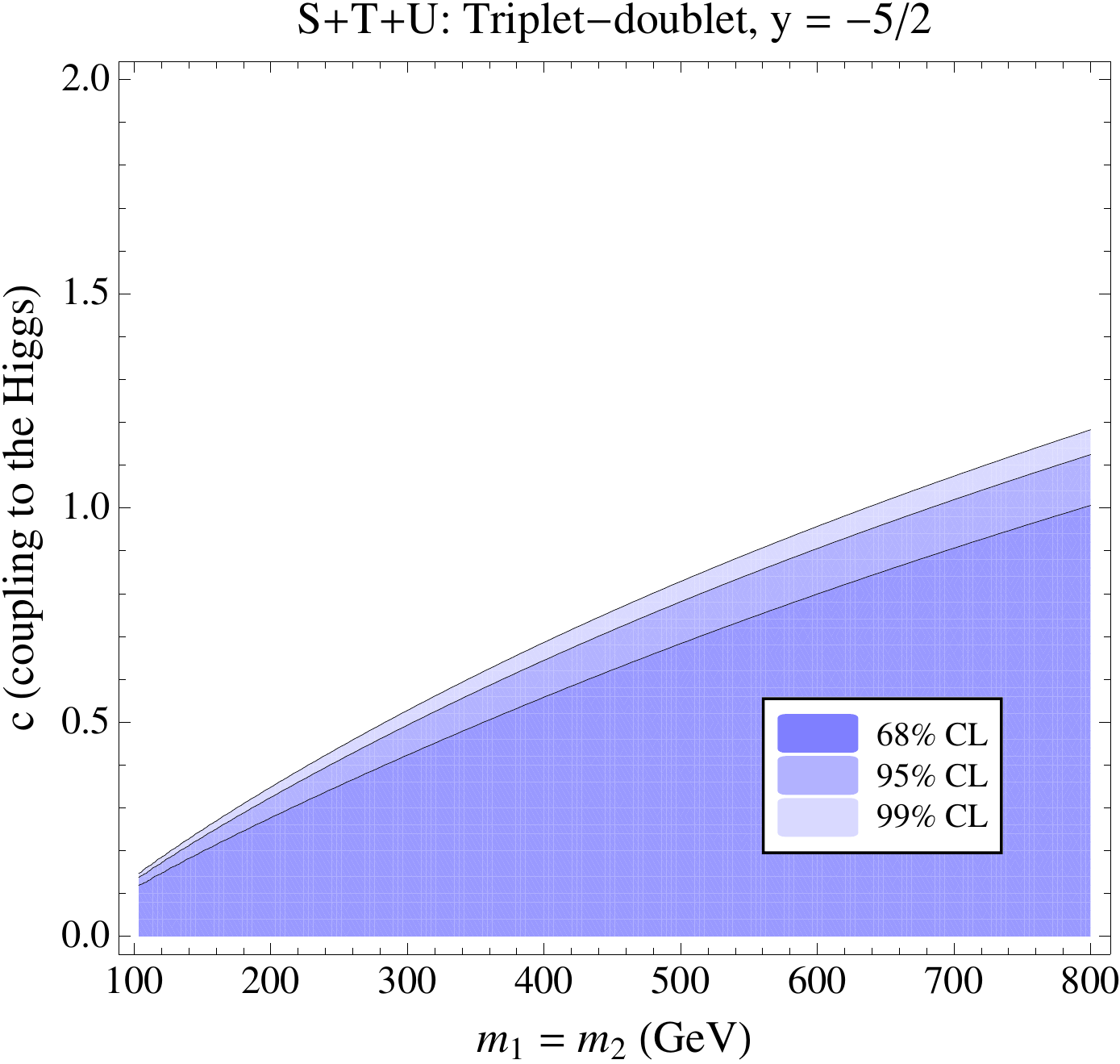}
 \includegraphics[width=0.49\textwidth]{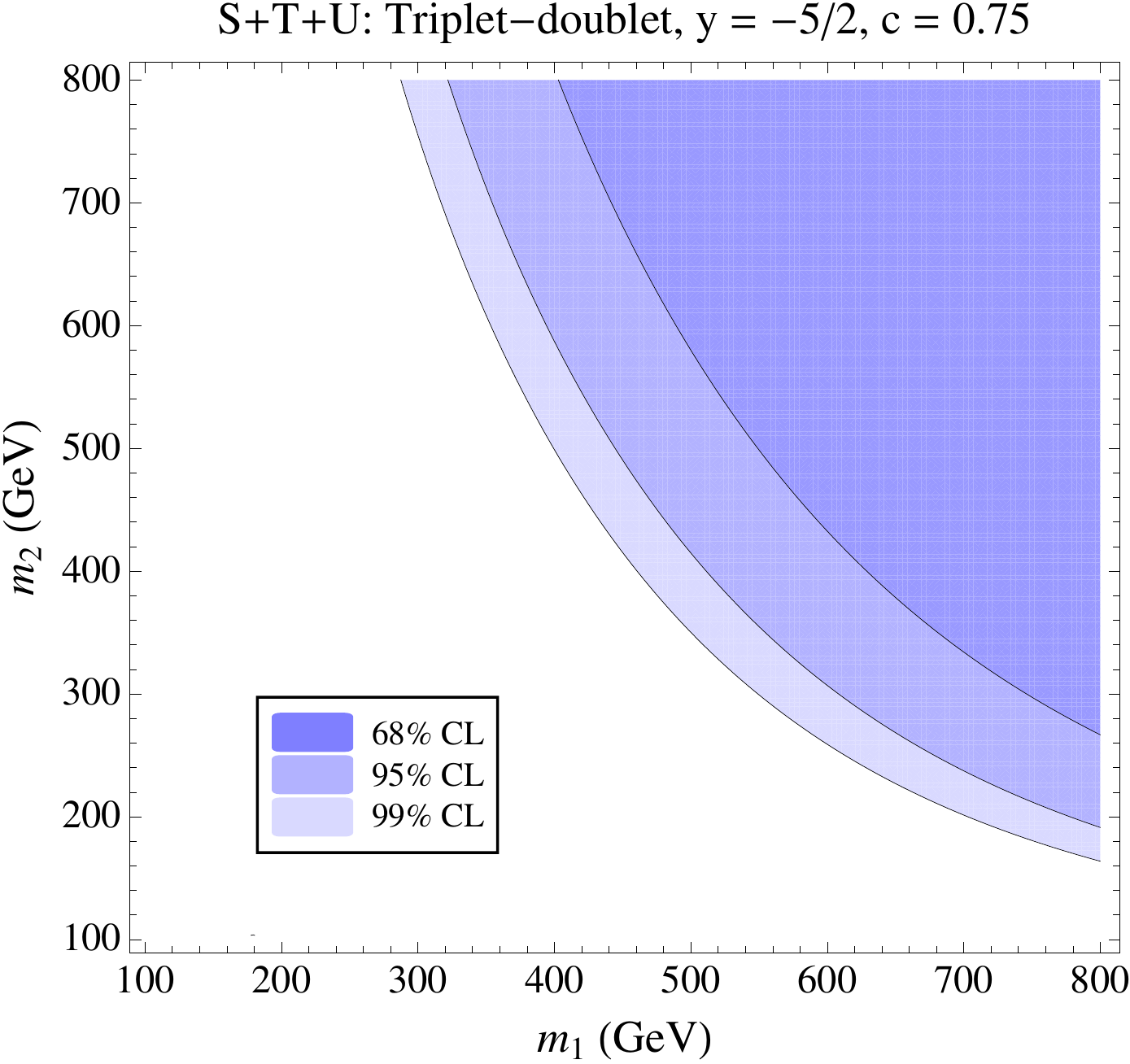}
\vspace{-2mm}
\end{center}
\vspace{-0.1cm}
\caption{Allowed region for the combined fit of S, T and U 
for $y=-5/2$ and $m_1=m_2$ (left panel) and 
for $y=-5/2$ and $c=0.75$ (right panel) in the triplet-doublet model.}
\label{fig:ST32-comb}
\end{figure*}

\begin{figure*}[!t]
\begin{center}
 \includegraphics[width=0.49\textwidth]{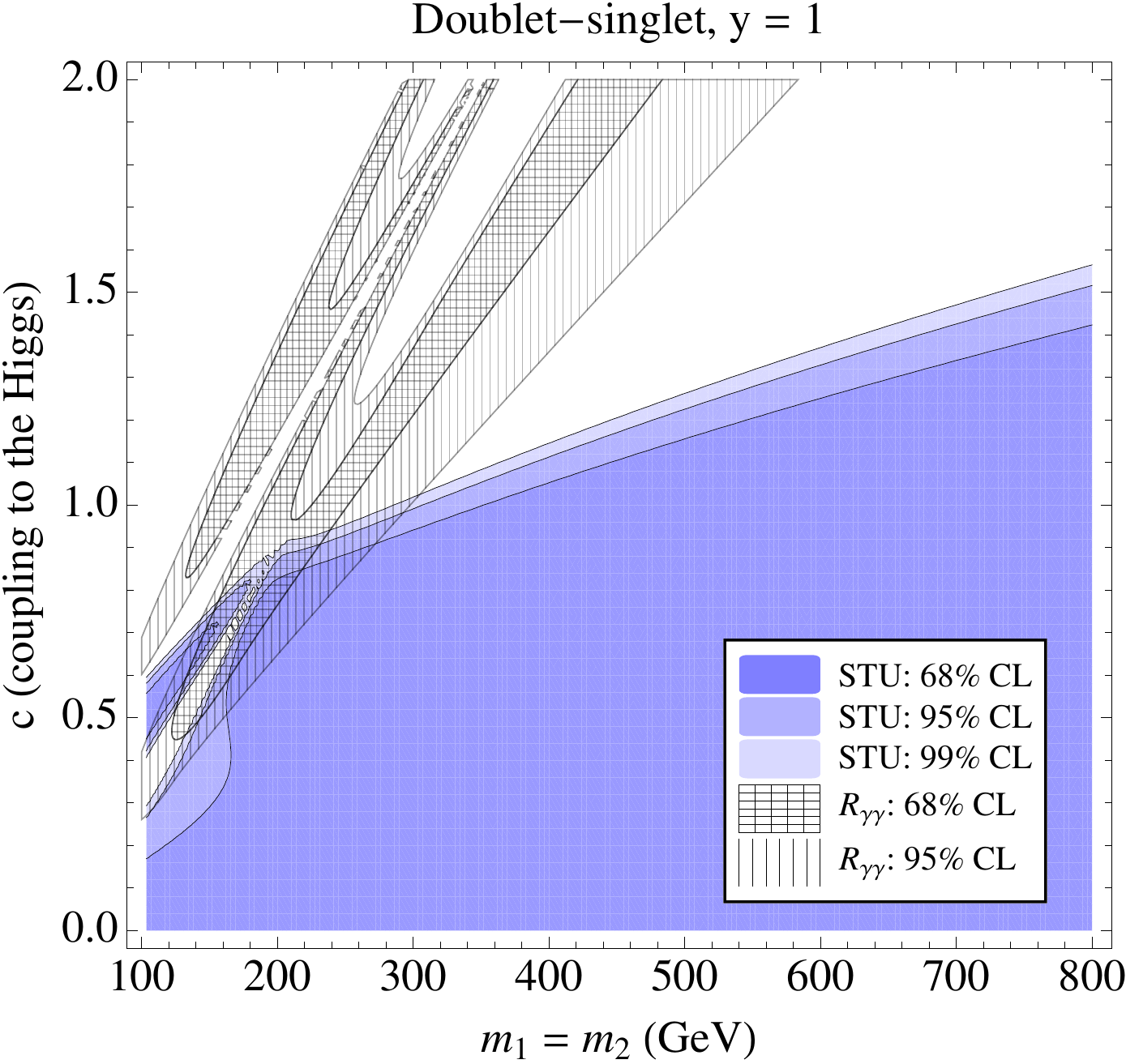}
\vspace{-2mm}
\end{center}
\vspace{-0.1cm}
\caption{Allowed region for the combined fit of S and T 
for $y=1$ in the doublet-singlet model
and its compatibility to $R_{\gamma\gamma}$.
We show the $R_{\gamma\gamma}$ allowed regions 
at 68\% and 95\% CL. }
\label{fig:TOT-2+1}
\end{figure*}

\begin{figure*}[!t]
\begin{center}
 \includegraphics[width=0.49\textwidth]{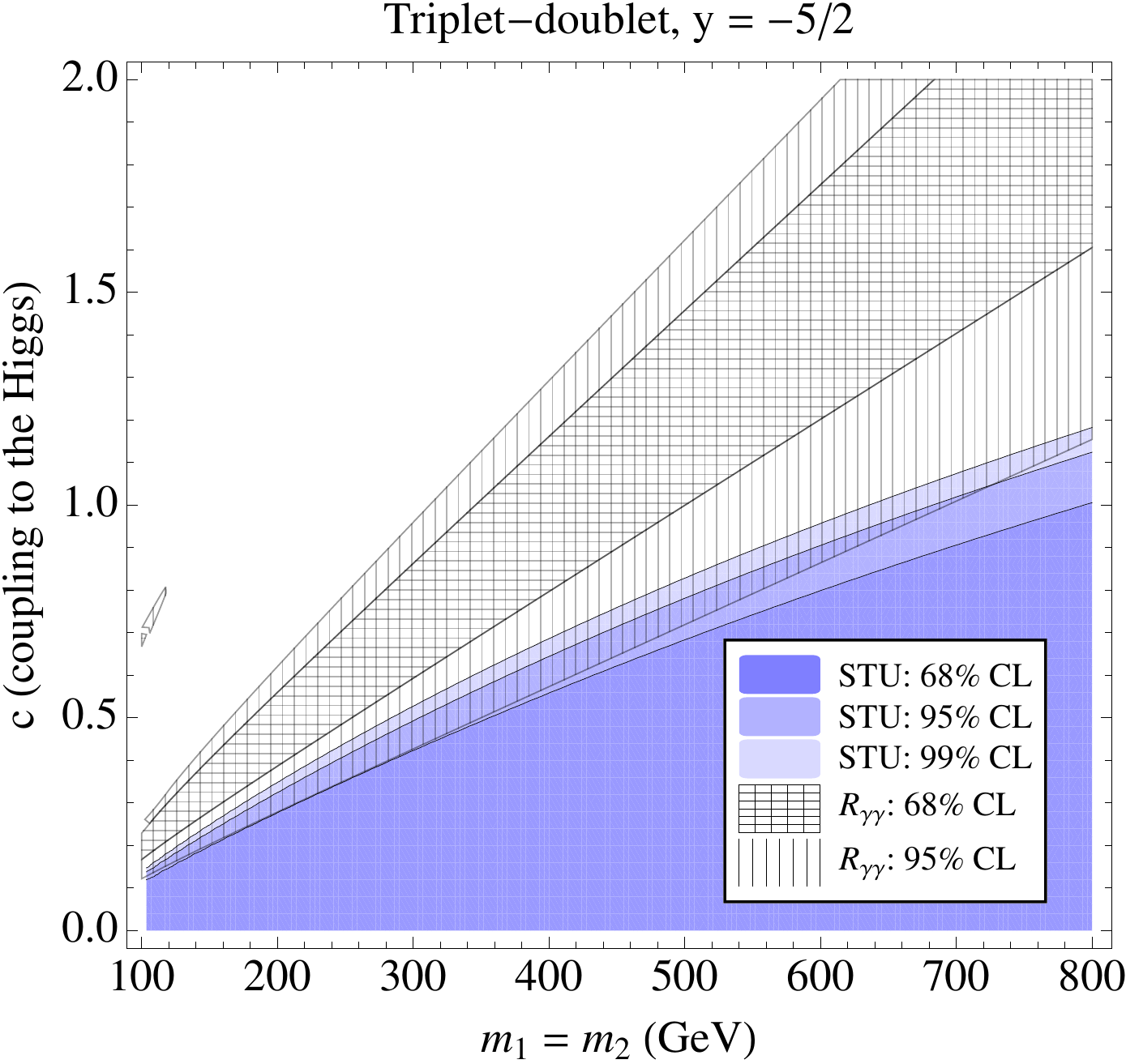}
\vspace{-2mm}
\end{center}
\vspace{-0.1cm}
\caption{Allowed region for the combined fit of S and T 
for $y=-5/2$ in the triplet-doublet model
and its compatibility to $R_{\gamma\gamma}$.
We show the $R_{\gamma\gamma}$ allowed regions 
at 68\% and 95\% CL. }
\label{fig:TOT-3+2}
\end{figure*}

\begin{figure*}[!t]
\begin{center}
\includegraphics[width=0.49\textwidth]{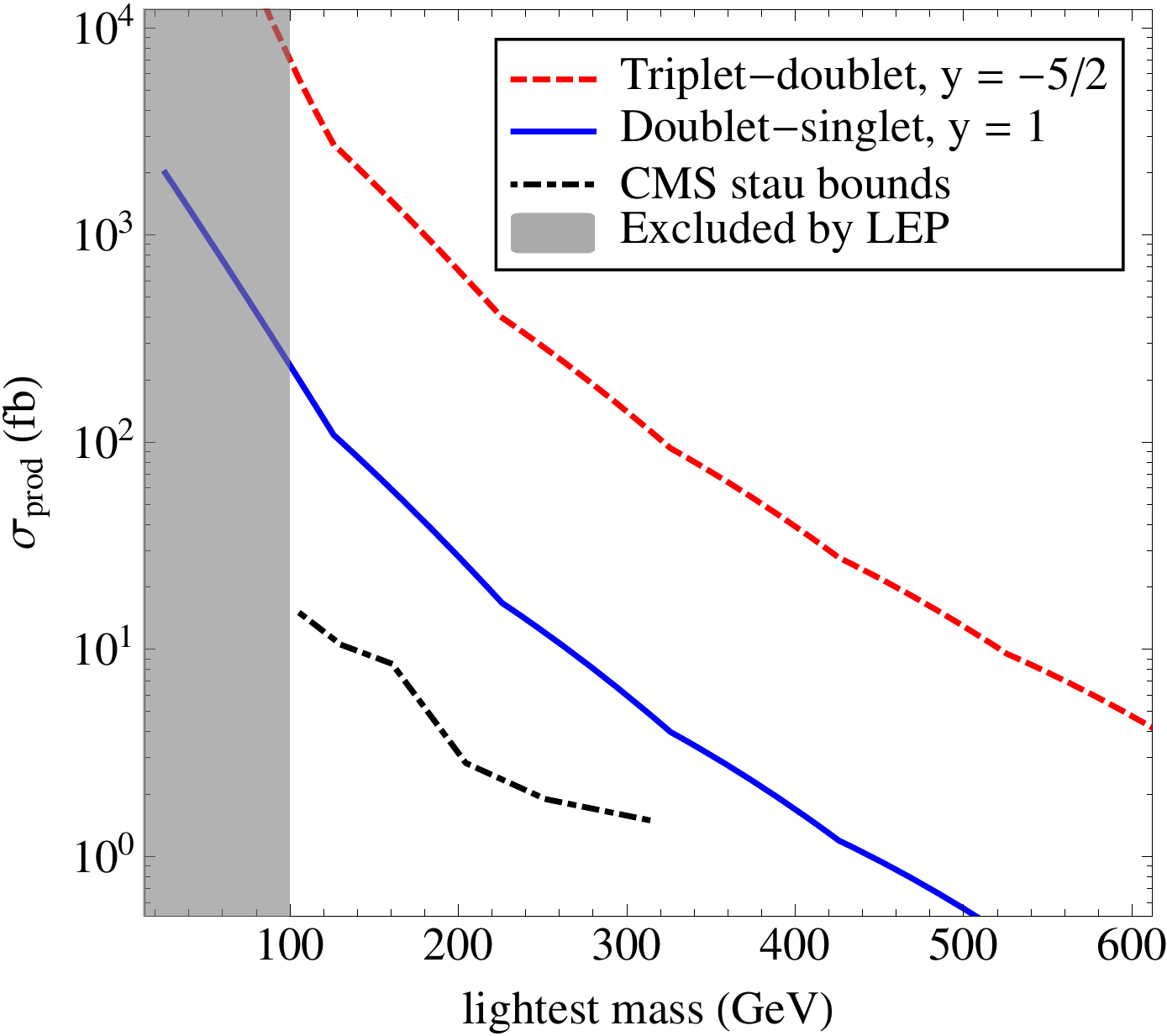}
\vspace{-2mm}
\end{center}
\vspace{-0.1cm}
\caption{Pair production cross section for the lightest new state 
in the doublet-singlet model with $y=1$ and in the 
triplet-doublet model with $y=-5/2$.
We also show the CMS limit for long lived staus pair 
production~\cite{stauspairproduction} at 7 TeV.}
\label{fig:crosspair}
\end{figure*}

\section{Conclusions}
\label{sec:conclusions}
We have investigated how the addition of extra uncolored fermion
states which do not mix to the Standard Model fermions, but do couple
to the Higgs, and therefore affects the $H\to \gamma \gamma$ rate.  We
focus on the two smallest possibilities for the SU(2)$_L$
representation of the fermion fields that can give rise to a
renormalizable lagrangian, coupling it to the Higgs, and contributing
to the Higgs diphoton width: the doublet-singlet model and the
triplet-double model.

We map the masses, charges and couplings that these possible states 
must have in order to satisfy the current limits on $\Gamma(H \to \gamma 
\gamma)$ from the LHC and the Tevatron Higgs combined 
data~\cite{Corbett:2012dm} and confront them with the 
allowed ones from electroweak precision measurements. 

We computed the pair production cross-section for the lightest states
and compared it to the current bounds from long-lived charged particles.

\begin{acknowledgments} 
  \vspace{-0.3cm} 
This work was supported by Funda\c{c}\~ao de Amparo
  \`a Pesquisa do Estado de S\~ao Paulo (FAPESP), Conselho Nacional 
de Desenvolvimento Cient\'ifico e Tecnol\'ogico (CNPq), by the European 
Commission under the contract PITN-GA-2009-237920 and by the Agence National 
de la Recherche under contract ANR 2010 BLANC 0413 01.  
We thank O.J.P. \'Eboli for useful comments and discussions.
\end{acknowledgments}

{\bf Note added:} During the finalization of this paper Ref.~\cite{cwagner,Nima} appeared.  
\appendix
\section{Appendix: Definitions of Some Functions Used in this Work}
Here we for completeness we define the loop functions used to compute  
$\Gamma(H \to \gamma \gamma)$.
\begin{eqnarray}
A_1(\tau) &=& \displaystyle -[2\tau^2 + 3 \tau + 3(2\tau-1) g(\tau)]/\tau^2 \\
A_{1/2}(\tau)&=& \displaystyle 2[\tau + (\tau-1) g(\tau)]/\tau^2\\
A_{0}(\tau)&=& \displaystyle -[\tau - g(\tau)]/\tau^2
\end{eqnarray}
where $g(\tau) = \arcsin^2 \sqrt{\tau}$, for $\tau \leq 1$. 

For the fermion gauge boson interaction lagrangian that can be generically 
written as 

$${\cal L}_{V f_i f_j}= \bar f_i (g^{ij}_{LV} P_L + g^{ij}_{RV}P_R) \gamma_\mu f_j 
V^\mu$$
we define the two point functions that enter in the oblique parameters 
calculation in terms of the generic couplings and of the universal 
functions $\Pi_{V \pm A}$ as~\cite{Dawson:2012di,hollik} 

\begin{equation}
\Pi_{V_1 V_2}(s) = (g^{ij}_{LV_1}g^{ij}_{LV_2} + g^{ij}_{RV_1}g^{ij}_{RV_2}) 
\, \Pi_{V+A}(s,m_i,m_j) +
(g^{ij}_{LV_1}g^{ij}_{RV_2} + g^{ij}_{LV_2}g^{ij}_{RV_1}) 
\, \Pi_{V-A}(s,m_i,m_j) ,
\label{2point-fermions}
\end{equation}
where a sum is implicit over all fermions involved, 
$V_1, V_2= W, Z, \gamma$, $m_i$ and $m_j$ are the masses of the 
fermions $f_i$ and $f_j$ in the loop and 

\begin{eqnarray}
\Pi_{V+A}(s,m_i,m_j)& =& \frac{N_c}{24 \pi^2} \left[ m_i^2\, \ln
  m_i^2 \left(1 - \frac{(m_i^2-m_j^2)}{2s} \right) + m_j^2\, \ln m_j^2
  \left(1 - \frac{(m_j^2-m_i^2)}{2s} \right) \right .\nonumber \\
&+&\left(s - \frac{(m_i^2+m_j^2)}{2} - \frac{(m_i^2-m_j^2)^2}{2s}
  \right) \left( \bar B_0(s,m_i,m_j) -\ln(m_i m_j)
  \right)  \nonumber \\
&-& \left .\frac{s}{3} +
  \frac{(m_i^2-m_j^2)^2}{2s} +\Delta_{\rm div} \right],
\label{piv+a}
\end{eqnarray}
and 
\begin{eqnarray}
\Pi_{V-A}(s,m_i,m_j)& =& \frac{N_c}{8 \pi^2} m_i m_j \left( \bar B_0(s,m_i,m_j) -\ln(m_i m_j) +\Delta_{\epsilon} \right),
\label{piv-a}
\end{eqnarray}
here $N_c$ are the number of colors, the divergent part 
$\Delta_{\rm div} \equiv \Delta_\epsilon (s-\frac{3}{2}(m_i^2+m_j^2))$
with $\Delta_\epsilon = \frac{2}{\epsilon}-\gamma+ \ln 4 \pi + \ln \mu^2$.

We have used the finite part of the $B_0$ function 
\begin{equation}
\bar B_0(s,m_i,m_j) = 1 -\frac{m_i^2+m_j^2}{m_i^2-m_j^2} \ln(\frac{m_i}{m_j})+F(s,m_i,m_j), 
\label{eq:b0finite}
\end{equation}
with 
\begin{equation}
F(s,m_i,m_j) = - 1 +\frac{m_i^2+m_j^2}{m_i^2-m_j^2}
\ln(\frac{m_i}{m_j}) - \int_0^1 dx \, \ln \left( \frac{x^2 -x
  (s+m_i^2-m_j^2)+m^2_i -i \epsilon}{m_i m_j}\right)\, .
\label{eq:f}
\end{equation}
as defined in Ref.~\cite{hollik}. Note that $F(0,m_i,m_j)=0$ so that $\bar B_0(0,m,m)=0$.

For $s=0$, the finite part of the previous expressions reads
\begin{equation}
 \Pi_{V+A}(0, m_i, m_j) = \frac{N_c}{24\pi^2} \frac{2 m_i^4(1-4\ln m_i) - 2 m_j^4(1-4\ln m_j)}{m_i^2-m_j^2},
\end{equation}
\begin{equation}
 \Pi_{V-A}(0, m_i, m_j) = \frac{N_c}{24\pi^2} \frac{8 m_i m_j \left( m_i^2 (2\ln m_i-1)- m_j^2 (2\ln m_j -1)\right)}{m_i^2-m_j^2}.
\end{equation}


\begin{thebibliography}{99}

\bibitem{ATLAS} ATLAS Collaboration, F. Gianotti, http://indico.cern.ch/conferenceDisplay.py?confId=197461.

\bibitem{CMS} CMS Collaboration, J. Incandela, http://indico.cern.ch/conferenceDisplay.py?confId=197461.


\bibitem{Corbett:2012dm} 
  T.~Corbett, O.~J.~P.~Eboli, J.~Gonzalez-Fraile and M.~C.~Gonzalez-Garcia,
  arXiv:1207.1344 [hep-ph].

\bibitem{Giardino:2012dp} 
  P.~P.~Giardino, K.~Kannike, M.~Raidal and A.~Strumia,
  arXiv:1207.1347 [hep-ph].

\bibitem{Espinosa:2012im} 
  J.~R.~Espinosa, C.~Grojean, M.~Muhlleitner and M.~Trott,
  arXiv:1207.1717 [hep-ph].

\bibitem{atlas7new}
ATLAS Collaboration, G.~Aad {\em et~al.},
\newblock (2012), arXiv:1207.0319.

\bibitem{Chatrchyan:2012tx}
CMS Collaboration, S.~Chatrchyan {\em et~al.},
\newblock Phys.Lett. {\bf B710}, 26 (2012), arXiv:1202.1488.

\bibitem{cmspashig12015}
CMS Collaboration,
\newblock CMS PAS HIG-12-015.

\bibitem{cmspashig12020}
CMS Collaboration,
\newblock CMS PAS HIG-12-020.

\bibitem{atlasconf12091}
ATLAS Collaboration,
\newblock ATLAS-CONF-2012-091.

\bibitem{Tevatron} The CDF Collaboration, the D0 Collaboration, 
the Tevatron New Physics, Higgs Working Group, arXiv:1207.0449.

\bibitem{Chatrchyan:2012tw} 
  S.~Chatrchyan {\it et al.}  [CMS Collaboration],
  Phys.\ Lett.\ B {\bf 710}, 403 (2012)
  [arXiv:1202.1487 [hep-ex]].


\bibitem{ATLAS:2012ad} 
  G.~Aad {\it et al.}  [ATLAS Collaboration],
  Phys.\ Rev.\ Lett.\  {\bf 108}, 111803 (2012)
  [arXiv:1202.1414 [hep-ex]].



\bibitem{Cao:2011pg} 
  J.~Cao, Z.~Heng, T.~Liu and J.~M.~Yang,
  Phys.\ Lett.\ B {\bf 703}, 462 (2011)
  [arXiv:1103.0631 [hep-ph]].

\bibitem{Alves:2011kc} 
  A.~Alves, E.~Ramirez Barreto, A.~G.~Dias, C.~A.~de S.Pires, F.~S.~Queiroz and P.~S.~Rodrigues da Silva,
  Phys.\ Rev.\ D {\bf 84}, 115004 (2011)
  [arXiv:1109.0238 [hep-ph]].

\bibitem{Carena:2011aa} 
  M.~Carena, S.~Gori, N.~R.~Shah and C.~E.~M.~Wagner,
  JHEP {\bf 1203}, 014 (2012)
  [arXiv:1112.3336 [hep-ph]].

\bibitem{Draper:2012xt} 
  P.~Draper and D.~McKeen,
  Phys.\ Rev.\ D {\bf 85}, 115023 (2012)
  [arXiv:1204.1061 [hep-ph]].

\bibitem{Kumar:2012ww} 
  K.~Kumar, R.~Vega-Morales and F.~Yu,
  arXiv:1205.4244 [hep-ph].


\bibitem{Dawson:2012di} 
  S.~Dawson and E.~Furlan,
  arXiv:1205.4733 [hep-ph].

\bibitem{Carena:2012gp} 
  M.~Carena, S.~Gori, N.~R.~Shah, C.~E.~M.~Wagner and L.~-T.~Wang,
  arXiv:1205.5842 [hep-ph].


\bibitem{Akeroyd:2012ms} 
  A.~G.~Akeroyd and S.~Moretti,
  arXiv:1206.0535 [hep-ph].


\bibitem{Carmi:2012zd} 
  D.~Carmi, A.~Falkowski, E.~Kuflik and T.~Volansky,
  arXiv:1206.4201 [hep-ph].

\bibitem{Carena:2012xa} 
  M.~Carena, I.~Low and C.~E.~M.~Wagner,
  arXiv:1206.1082 [hep-ph].


\bibitem{Carmi:2012in} 
  D.~Carmi, A.~Falkowski, E.~Kuflik, T.~Volansky and J.~Zupan,
  arXiv:1207.1718 [hep-ph].


\bibitem{Chang:2012ta} 
  W.~-F.~Chang, J.~N.~Ng and J.~M.~S.~Wu,
  arXiv:1206.5047 [hep-ph].

\bibitem{Bertolini:2012gu} 
  D.~Bertolini and M.~McCullough,
  arXiv:1207.4209 [hep-ph].


\bibitem{Altarelli:1990zd} 
  G.~Altarelli and R.~Barbieri,
  Phys.\ Lett.\ B {\bf 253}, 161 (1991).

\bibitem{Peskin:1991sw} 
  M.~E.~Peskin and T.~Takeuchi,
  Phys.\ Rev.\ D {\bf 46}, 381 (1992).


\bibitem{Baak:2011ze} 
M.~Baak, M.~Goebel, J.~Haller, A.~Hoecker,
  D.~Ludwig, K.~Moenig, M.~Schott and J.~Stelzer,
  Eur.\ Phys.\ J.\ C {\bf 72}, 2003 (2012)
  [arXiv:1107.0975 [hep-ph]].


\bibitem{calchep} A. Pukhov, CalcHEP 2.3: MSSM, structure functions,
  event generation, batchs, and generation of matrix elements for
  other packages, arXiv:hep-ph/0412191.

\bibitem{CMS4th}
CMS collaboration, PAS EXO-11-098


\bibitem{Achard:2001qw} 
  P.~Achard {\it et al.}  [L3 Collaboration],
  Phys.\ Lett.\ B {\bf 517}, 75 (2001)
  [hep-ex/0107015].


\bibitem{stauspairproduction} S.~Chatrchyan {\it et al.}  [CMS
  Collaboration],
  Phys.\ Lett.\ B {\bf 713}, 408 (2012)
  [arXiv:1205.0272 [hep-ex]].

\bibitem{Pumplin:2002vw} 
  J.~Pumplin, D.~R.~Stump, J.~Huston, H.~L.~Lai, P.~M.~Nadolsky and W.~K.~Tung,
  JHEP {\bf 0207}, 012 (2002)
  [hep-ph/0201195].


\bibitem{cwagner} A. Joglekar, P. Schwaller and C. E. M. Wagner,
arXiv:1207.4235.

\bibitem{Nima} N. Arkani-Hamed, K. Blum, R. Tito D'Agnolo and 
J. Fan, arXiv:1207.4482.

\bibitem{hollik} W.F.L. Hollik, DESY 88-188. 

\end{thebibliography}
\end{document}